\documentclass{aastex}
\usepackage{natbib}
\shorttitle{Compositional Diversity of Extrasolar Terrestrial Planets}
\shortauthors{Bond et al.}

\begin{document}

\title{The Compositional Diversity of Extrasolar Terrestrial Planets: I. In$-$Situ Simulations}

\author{Jade C. Bond\altaffilmark{1,3}}
\affil{Lunar and Planetary Laboratory, University of Arizona,
    Tucson, AZ 85721}
\email{jbond@psi.edu}
\and
\author{David P. O'Brien\altaffilmark{2}}
\affil{Planetary Science Institute, Tucson, AZ 85719}
\and
\author{Dante S. Lauretta\altaffilmark{1}}
\affil{Lunar and Planetary Laboratory, University of Arizona,
    Tucson, AZ 85721}

\altaffiltext{1}{Lunar and Planetary Laboratory, University of Arizona, 1629 East University Boulevard, Tucson, AZ 85721-0092.}
\altaffiltext{2}{Planetary Science Institute, 1700 E. Fort Lowell, Tucson, AZ 85719.}
\altaffiltext{3}{Now at Planetary Science Institute.}

\begin{abstract}
Extrasolar planet host stars have been found to be enriched in key
planet-building elements. These enrichments have the potential to drastically
alter the composition of material available for terrestrial planet
formation. Here we report on the combination of dynamical models of late-stage terrestrial planet formation within
known extrasolar planetary systems with chemical equilibrium models of the
composition of solid material within the disk. This allows us to determine the
bulk elemental composition of simulated extrasolar terrestrial planets. A wide
variety of resulting planetary compositions are found, ranging from those that
are essentially "Earth-like", containing metallic Fe and Mg-silicates, to those
that are dominated by graphite and SiC. This shows that a diverse range
of terrestrial planets may exist within extrasolar planetary systems.
\end{abstract}

\keywords{planets and satellites: composition --- planets and satellites: formation --- planetary systems}

\section{Introduction}
Extrasolar terrestrial planets are a tantalizing prospect. Given that the number
of planets in the galaxy is expected to correlate inversely with planetary mass,
it is expected that Earth-sized terrestrial planets are much more common than
giant planets \citep{marcy2000}. Although still undetectable by current
exoplanet searches, the possibility of their existence in extrasolar planetary
systems has been examined by several authors. Many such studies have focussed on
the long term dynamical stability of regions within the planetary system where
such planets could exist for geologic timescales \citep{br1,br2,asg}. Several
systems have been found to posses such regions (e.g. \citealt{br1}), indicating
that if they are able to form, terrestrial planets may still be present within
extrasolar planetary systems. Analyses of this nature are of great interest to future planet search
missions as they assist in constraining future planet search targets. However,
they provide little insight into the formation mechanism and physical
and chemical properties of such planets and do not necessarily indicate the
presence of a terrestrial planetary companion.

A few other studies have gone one step further and undertaken detailed
N-body simulations of terrestrial planet formation. \cite{ray05} considered terrestrial planet formation in a series of
hypothetical `hot Jupiter' simulations and found that terrestrial planets can
indeed form in such systems (beyond the orbit of the giant planet) provided the `hot Jupiter' is located within 0.5AU
from the host star. Furthermore, such planets may even have water contents
comparable to that of the Earth. Terrestrial planets have been found to form even
in simulations of systems which have undergone large-scale migration of the giant planet
\citep{raymond:2006,avi}. Terrestrial planets were found to form both exterior and interior
to the giant planet after migration has occurred and many were located within
the habitable zone of the host star. As many extrasolar planets are believed to
have experienced such a migration, it is encouraging that terrestrial planets may still be able to form within these systems. To date, only \cite{br3} have
undertaken terrestrial planet formation simulations for specific planetary
systems. They considered four known planetary systems and found that terrestrial
planets could form in one of the systems (55Cancri). Small bodies comparable to
asteroid sized objects would be stable in another (HD38529).

An even more intriguing question beyond whether or not terrestrial planets could exist within these systems is their potential chemical
composition. Extrasolar planetary host stars are already known to be chemically
unusual \citep{g1,g2,butler00,g4,g5,g3,sb,s1,ism04,g6,sm,re,fv,bond1,bond:2008},
displaying systematic enrichments in Fe and smaller, less statistically
significant enrichments in other species such as C, Si, Mg and Al
\citep{g3,g6,sb,bo,fv,be,bond1}. Given that these enrichments are likely primordial in
origin \citep{s1,s2,s04,s05,fv,bond1}, it is thus likely that the
planet forming material within these systems will be similarly enriched. Hints
of such a correlation between transiting giant planets and stellar metallicity
have been observed \citep{metal1,metal2}. Consequently, it is likely that
terrestrial extrasolar planets may have compositions reflecting the enrichments
observed in the host stars. Furthermore, several known host stars have been
found to have C/O values above 0.8 \citep{bond:2008}. Systems with high C/O
ratios will contain large amounts of C phases (such as SiC, TiC and graphite),
resulting in any terrestrial planets within these systems being enriched in C
and potentially having compositions and mineralogies unlike any body observed within our Solar System.

Despite the likely chemical peculiarities of extrasolar planetary systems and the early successes of extrasolar terrestrial
planet formation simulations, no studies of extrasolar terrestrial planet
formation completed to date have considered both the dynamics of formation and
the detailed chemical compositions of the final terrestrial planets produced.
This study addresses this issue by simulating
late-stage in-situ terrestrial planet formation within ten extrasolar planetary
systems while simultaneously determining the bulk elemental compositions of the
planets produced. This is the first such study to consider both the dynamical
and chemical nature of potential extrasolar terrestrial planets and it
represents a significant step towards understanding the diversity of potential extrasolar
terrestrial planets.

\section{System Composition}
The two most important elemental ratios for determining the mineralogy of
extrasolar terrestrial planets are C/O and Mg/Si. Note that throughout this paper, Mg/Si and C/O refers to the elemental number ratios, \emph{not} solar normalized logarithmic values often quoted in stellar spectroscopy (usually shown as [X/H] for the solar normalized logarithmic abundance of element X compared to H). The ratio of C/O controls the
distribution of Si among carbide and oxide species. Under the assumption of
equilibrium, if the C/O ratio is greater than 0.8 (for a pressure of 10$^{-4}$
bar), Si exists in solid form primarily as SiC. Additionally, a
significant amount of solid C is also present as a
planet building material. For C/O values below 0.8, Si is present in
rock-forming minerals as SiO$_{4}$$^{4-}$ (or SiO$_{2}$), allowing for the formation of silicates. The
silicate mineralogy is controlled by the Mg/Si value. For Mg/Si values less than
1, Mg is in pyroxene (MgSiO$_{3}$) and the excess Si is present as other
silicate species such as feldspars. For Mg/Si values ranging from 1 to 2, Mg is
distributed between olivine (Mg$_{2}$SiO$_{4}$) and pyroxene. For Mg/Si values
extending beyond 2, all available Si is consumed to form olivine with excess Mg
available to bond with other elements as MgO or MgS.

Just as stellar C/O values are known to vary within the solar neighborhood
\citep{Gustafsson:99}, the C/O values of extrasolar
planetary systems also deviate from the solar value. The photospheric C/O
vs. Mg/Si values for known extrasolar planetary host stars are shown in Figure
\ref{regions}, based on stellar abundances taken from \cite{gilli} (Si and Mg),
\cite{be} (Mg), \cite{eca} (C) and \cite{oxygen} (O). A conservative approach
was taken in determining the average error shown in Figure \ref{regions}. The
errors published for each elemental abundance were taken as being the 2$\sigma$
errors (as the method used to determine them naturally provides the 2$\sigma$ error range) and were used to determine the maximum and minimum abundance values
possible with 2$\sigma$ confidence for each system. The elemental ratios
produced by these extremum abundances were thus taken as the 2$\sigma$ range in
ratio values and are shown as errors in Figure \ref{regions}.

The mean values of Mg/Si and C/O for all extrasolar planetary systems for which
reliable abundances are available are 1.32 and 0.77 respectively, which are
above solar values (Mg/Si$_{\bigodot}$ = 1.00 and C/O$_{\bigodot}$ = 0.54)
\citep{asp}. This non-solar average and observed variation implies that a wide
variety of materials would be available to build terrestrial planets in
those systems, and not all planets that form can be expected to be similar to
that of Earth. Of the 60 systems shown, 21 have C/O values above 0.8, implying
that carbide minerals are important planet building materials in potentially more than 30\%
of known planetary systems. This implies that a similar fraction of
protoplanetary disks should contain high abundances of carbonaceous grains.
As comets represent some of the most primitive material within our planetary system, it is likely that a similar mass fraction will apply to the protoplanetary nebula. Furthermore, infrared
spectral features at 3.43 and 3.53 $\rm \mu$m observed in 4\% of protoplanetary disks have been identified as being produced by nano-diamonds \citep{acke}. Such high
abundances of carbon-rich grains in nascent planetary systems is inconceivable
if they have primary mineralogy similar to our Solar System, thus implying that
C-rich planetary systems may be more common than previously thought. The idea of
C-rich planets is not new \citep{kuchner} but the potential prevalence of these bodies has
not been previously recognized, nor have specific systems been identified as
likely C-rich planetary hosts. These data clearly demonstrate that there are a significant number of systems in
which terrestrial planets could have compositions vastly different to
any body observed in our Solar System.

Both host and non-host stars\footnote{Throughout this paper, non-host stars refers to stars observed as part of a planet search program that are not currently known to harbor a planetary companion.} display the same distributions in C/O and Mg/Si
values (see Figure \ref{sicomp1}). The mean, median and standard deviation for
both the host and non-host stars is shown in Table \ref{compsi}. The values
listed in Table \ref{compsi} are based on the stellar abundances determined in
\cite{bond:2008} as \cite{gilli,be,oxygen,eca} provide abundances for all four
elements for just three non-host stars, thus preventing host and non-host
comparisons. It is essential to point out here that the values shown in Figure \ref{sicomp1} and Table \ref{compsi} are based on a different dataset than is used for the simulations presented in this paper and are meant for comparative purposes only. The [Mg/H] values in \cite{bond:2008} are known to be lower than for previously published common stars and the different spectral indicators were used to obtain O abundances (see \cite{bond:2008} for more details). However, as both host and non-host stars were examined in the same way, the use of this data to compare the two populations is still valid. Given the excellent agreement between host and non-host stars, we
conclude that known planetary host stars are not preferentially biased towards
higher C/O or Mg/Si values compared to stars not known to harbor a planetary
companion. This in turn implies that the prevalence of C-rich planetary systems
identified above is not statistically unusual (in terms of stellar composition).

However, a high degree of uncertainty is associated with all stellar C/O and
Mg/Si values. The primary source of this error is uncertainty in the stellar
elemental abundances themselves. Spectrally determined abundances are sensitive
to continuum placement and stellar atmospheric parameters such as stellar effective temperature (T$_{\rm eff}$), stellar surface gravity (log g), metallicity and microturbulence. These
uncertainties result in an average 2-$\sigma$ error of $\pm$0.04 for [Mg/H], $\pm$0.07 for
[Si/H], $\pm$0.08 for [C/H] and $\pm$0.09 for [O/H] \footnote{[X/H] = log(X/H)$_{\star}$ - log(X/H)$_{\odot}$ where X/H is the ratio of the abundance of a given element, X, to H.}. These
errors result in considerable percentage uncertainties for the Mg/Si and C/O
values of up to 124\%. Although large, the errors will not decrease until we are
able to reduce the uncertainty on each individual stellar elemental
abundance.

Of the four key elements discussed thus far, O is by far the most controversial.
Three different spectral lines are available for determining the photospheric O
abundance - the forbidden OI lines located at 6300 and 6363 \AA, the OI triplet
located between 7771 \AA\ and 7775 \AA\ and the OH lines located near 3100 \AA\
\citep{oxygen}. Previous studies have found discrepancies between abundances
obtained from different spectral lines for the same star of up to 1 dex \citep{iserror} (A dex is the logarithmic unit for elemental abundances). Each of these lines is subject to interferences from different
stellar sources. The forbidden OI lines are weak and blended with Ni, the OI
triplet is influenced by non-local thermodynamic equilibrium (non-LTE) effects
and the OH lines are influenced by stellar surface features \citep{oxygen}.
\cite{oxygen} undertook a detailed examination of the correlation between these
different O indicators for a sample of 96 host and 59 non-host stars. Their
study showed that the discrepancies in abundances determined by the three
different indicators was less than 0.2dex for the majority of stars examined.
The forbidden OI and OH lines were found to be in good agreement with each other
while abundances obtained from the O triplet lines (with the appropriate NLTE
corrections applied to them) were systematically lower. However, all three
indicators produced abundances in keeping with the galactic evolutionary trends
observed for lower metallicity (and thus younger) stars \citep{oxygen}. The C/O
ratios shown in Figure \ref{regions} (and used throughout this study) are based
on the O abundances from \cite{oxygen} obtained from the forbidden OI spectral
line observed at 6300.3 \AA\ as abundances from this line are in agreement with
the abundances obtained from the OH line and produce a marginally better fit to
stellar evolutionary models.

The solar O abundance itself has experienced a `crisis' in recent years
\citep{ayres} with several studies suggesting that a downward revision of the
solar O value is required (e.g. \citealt{ayres,socas}). However, realistic
errors for the suggested new O abundance are approximately 0.1 dex \citep{socas}
and the abundances of the present study vary from [O/H] = 0.38 (HD 177830) to
[O/H] = $-$0.10 (HD 108874), a range of 0.48 dex. As our present study range is
almost five times larger than the errors of the revised O abundance, we feel
confident in the compositional variations identified here as being caused by
variations in the stellar O abundances of specific systems.

Given the errors associated with each individual elemental abundance (and thus
also ratio value), it is natural to consider the uncertainty associated
with the dispersion seen in Figure \ref{regions}. The observed dispersion is
produced by both dispersion in the actual values and dispersion due to
measurement errors. Propagating the errors of the individual abundances leads to a standard deviation of 0.22, slightly reduced from 0.27 without considering errors. Thus it is probable that the real range in
elemental ratios is less than is shown in Figure \ref{regions} and fewer
planetary systems have C/O values above 0.8. Based on the errors described above
and shown in Figure \ref{regions}, only 5 of the 61 planetary systems shown (8\%
of the sample) can be said with 2$\sigma$ confidence to have C/O values above
0.8. This increases to just 7 planetary systems (11\% of the sample) when we
reduce the confidence interval to 1$\sigma$. It should also be noted here that the O abundances utilized here are on average higher than those produced by other indicators. As such, the C/O values are lower and should be considered conservative values. Additionally, as the errors are bidirectional, it is also possible that our high C/O sample may be larger. Making the overly-optimistic assumption that all systems lie at the upper bounds of their error range, 47 of the 61 planetary systems (77\% of the sample) may have C/O values above 0.8.

Taken together, this suggests that while the number of systems with C/O values larger than 0.8 may be less than indicated in Figure \ref{regions}, it is likely not as low as the most conservative estimates, and a significant number of planetary systems may have C-rich condensation sequences. Planets that form in these systems would contain significant amounts of carbide phases as major planet building materials, and would differ significantly from the silicate-dominated planets that form in systems with lower C/O values.

\section{Simulations}
\subsection{Extrasolar Planetary Systems}
Ten known extrasolar planetary systems spanning the entire
compositional spectrum of observed planetary systems were selected for
this study. This wide range was purposefully chosen to explore the
full diversity of possible extrasolar terrestrial planets. The dynamical and
chemical characteristics of each system are shown in Tables
\ref{EGPprop}, (orbital parameters of known planetary companions), \ref{EGPchem}
(stellar abundances in logarithmic units) and \ref{EGPchemSi} (stellar
abundances normalized to 10$^{6}$ Si atoms), while the known giant planet
architecture is shown in Figure \ref{all_sys} and the C/O and Mg/Si values were previously
shown in Figure \ref{regions}. Dynamical properties of the known giant planets
were taken from the catalog of \cite{but-cat} [these values were
augmented with more recent values from http://exoplanets.org when available]. Note
that the innermost planet of 55Cnc was neglected in our present simulations due
to its location and low mass.

\subsection{Chemical Simulations}
As in \cite{bond:2009}, the chemical composition of material within the disk is
assumed to be determined by equilibrium condensation within the primordial
stellar nebula. Although the timescales for formation of planetesimals and embryos considered here can be comparable to the timescale for reaction kinetics (eg. less than 1$\times$10$^{6}$yr for the Jovian circumplanetary disk \citep{mousis:2006}), the hot inner mid-plane region (<10AU from the host star) achieves thermodynamic equilibrium within approximately 100 to 10,000 years \citep{semenov:2010}. Therefore the assumption that the simulated planetesimals form in equilibrium with their surroundings is still valid for this type of study. This approach is further supported by evidence observed within our own Solar System. Primitive chondritic material has been found to be remarkably similar to that predicted by equilibrium condensation studies with a Solar composition \citep{ebel}, with bulk elemental abundances that are also functions of their equilibrium condensation temperature for relevant disk conditions \citep{davis}. Preserved equilibrium compositions can still be seen today in the observed thermal stratification of the asteroid belt \citep{gradie}. Although some uncertainty as to the precise cause of these features still lingers, the fact remains that we can utilize equilibrium condensation temperatures to determine the bulk elemental abundances of solid material within planetary systems.

Equilibrium condensation sequences for an identical list of
elements as used in \cite{bond:2009} (H, He, C, N, O, Na, Mg, Al, Si, P, S, Ca,
Ti, Cr, Fe and Ni) were obtained using the commercial software package HSC
Chemistry (v. 5.1) using the same list of gaseous and solid species as in
\cite{bond:2009} (listed here in Table \ref{HSC}). As these elements represent the major solid-forming species, no significant limitations are encountered by neglecting other species from these simulations.

Observed stellar photospheric abundances were adopted as a proxy for the
composition of the stellar nebula and were taken from \cite{gilli,be,ec,oxygen}.
It should be noted that all abundances applied here were taken from the same
research group and were obtained from the same spectra thus acting to limit any
possible systematic differences in abundances due to instrument or
methodological differences between various studies. The input values used in HSC
Chemistry for each system are shown in Table \ref{inputs}. All species are
initially assumed to be in their elemental and gaseous form and no other species
or elements are assumed to be present within the system.

Neither N or P abundances have been obtained for extrasolar planetary host
stars, primarily due to the difficulty in finding unblended spectral lines to
use within the visual spectral range (where most studies have been focussed).
For this present study, we overcame this issue by obtaining approximate
abundances for both elements based on the well known odd-even effect. Caused by
the increased stability of even atomic number nuclei relative to odd-numbered
nuclei, this effect produces the observed sawtooth pattern in the Solar
elemental abundances. Extrasolar abundances were obtained by fitting a linear trend through the solar abundances for the same elements studied here and then applying this same fit to observed extrasolar host star
abundances of Na and Al. This approach assumes that extrasolar host stars will
display the same atomic sawtooth pattern, a valid assumption as host stars do
not appear to have undergone any form of systematic processing (such as
pollution by a nearby supernova event) to cause a significant deviation
\citep[see][]{bond:2008}. The same approach was adopted for those systems without an observed [S/H] value.

As for the Solar System simulations of \cite{bond:2009}, ``nominal'' radial pressure and temperature profiles obtained from the \cite{hersant}
protoplanetary disk midplane model  were used to correlate chemical compositions with a spatial location within the disk. The stellar mass accretion rate $\dot{M}$, a primary input to the Hersant model, is known
to vary with stellar mass as:

\begin{equation}
{\rm \dot{M}}\:\propto\:\rm M^{3/2}
\end{equation}\\

\noindent where M is the mass of the star in solar masses (ranging from 0.98 to 1.48M$_{\bigodot}$) and $\dot{M}$ is the
mass accretion rate in solar masses per year. The input mass accretion rates were thus scaled for the
stellar mass of the host star obtained from the Simbad
database\footnote{accessed at http://simbad.u-strasbg.fr/simbad/}. The resulting
stellar accretion rates obtained are shown in Table \ref{mdot}. All other input
parameters for the \cite{hersant} models remained unchanged ($\rm \alpha$ = 0.009, initial disk radius = 17AU). Minor differences in temperature and pressure profiles were produced for the various systems simulated. For example, the temperature and pressure values at a distance of 1AU from the host star for the systems with the highest and lowest mass accretion rates differed by just 51K and 1.6$\times$10$^{-6}$ bars respectively.

It is quite possible that the mass accretion rate varies throughout the disk itself, potentially producing variations in midplane pressure and temperature values. The \cite{hersant} models assumed a uniform mass accretion rate. However, the conditions at 1AU in the system with the highest mass accretion rate (HD177830) occur just 0.09AU closer to the host star in the system with the lowest mass accretion rate (HD72659). This indicates that solid material will still condense out at essentially the same radial location, despite variations in the mass accretion rate of the disk. As such, any modification of midplane conditions from those used here due to a varying mass accretion rate are likely to be small enough as to be neglected for the purposes of this study. It is also important to note that the current
approach does not include variations in the midplane conditions produced by
different chemical compositions (which would alter parameters such as disk
viscosity and opacity), nor does it include the effects of stellar
luminosity (which aren't incorporated into the \cite{hersant} model). As such,
the scaling applied here is a somewhat simplistic approach to a complex issue but is
valid for the current aims of this study.

Midplane pressure and temperature values were determined with an average radial
separation of 0.01AU throughout the study region. As in \cite{bond:2009}, an
ensemble of disk compositions was determined based on \cite{hersant} disk
conditions at seven different evolutionary times (t = 2.5$\times$10$^{5}$yr,
5$\times$10$^{5}$yr, 1$\times$10$^{6}$yr, 1.5$\times$10$^{6}$yr,
2$\times$10$^{6}$yr, 2.5$\times$10$^{6}$yr, 3$\times$10$^{6}$yr). However, as we
previously found the best fit to known planetary values in the Solar System (specifically the compositions of Venus, Earth and Mars) to
occur using disk conditions obtained for t = 5$\times$10$^{5}$yr
\citep{bond:2009}, our discussions will mostly focus on compositions produced by
disk conditions at this time. Note that these times only refer to the timing of the thermodynamical conditions under which the planetesimals condensed within the disk.

\subsection{Dynamical Simulations\label{dyn}}
In this study, we build upon our recent success in simulating
terrestrial planet formation within the Solar System \citep{bond:2009} and apply
the same methodology here. Terrestrial planet formation in extrasolar
planetary systems is a complex problem, especially when giant planet migration
is taken into account, and modeling it in detail for even a single system is
computationally expensive \citep{raymond:2006,avi}. In the interest of exploring a wide
range of systems, we use a more basic approach here which necessarily neglects
some of the complexities of planet formation, but still allow us to demonstrate
some of the potential implications of varying system compositions on final
terrestrial planet compositions. More detailed dynamical simulations will be
the subject of future work.

N-body simulations of terrestrial planet accretion in each of the selected
extrasolar planetary systems are run using the SyMBA n-body integrator
\citep{duncan}. The orbital parameters of the giant planets in each system are
taken from the catalog of \cite{but-cat}, and updated with values from
exoplanets.org as additional data on these systems were obtained. All values utilized are shown in Table \ref{EGPprop}. Inclinations
of each of the giant planets are assumed to be zero since no such measurements
have been obtained for these systems. Due to the computational time required,
current simulations only contain an initial population of roughly Lunar-
to Mars-mass embryos (i.e. no planetesimal swarm is included). This differs
from the Solar System simulations of \cite{dave} as used in
\cite{bond:2009} as those simulations included a planetesimal swarm. As such, it is possible that differences between the dynamical simulations performed here and those of the Solar System as used in
\cite{bond:2009} will occur, most likely in that the dynamical
excitation of the resulting systems system will be somewhat larger than they
would be if a planetesimal swarm were present.

For each extrasolar planetary system modeled, planetary embryos are distributed
in the zone between the star and the giant planet (or in the case where there
are one or more giant planets close to the host star such as 55Cnc and
Gl777, in the region between the inner and outer giant planets) according to
the relations between embryo mass, spacing, and orbital radius given by
\cite{kandi}. No embryos are initially located interior to 0.3 AU. The timestep
for the integration was set to at least 20$\times$ the orbital period of the
innermost planet or planetary embryo, or the orbital period of a body at 0.1 AU,
whichever is smaller (this corresponds to a half-day timestep for an
inner radius of $~$0.1 AU), and the simulations are run for $~$100-250 Myr.
Surface mass density profiles that vary as r$^{-3/2}$, normalized to 10
gcm$^{-2}$ at 1 AU, were assumed. For each system, a
minimum of 4 accretion simulations were run, using different random
number seeds for generating the initial embryo distribution.

Migration of the giant planets is very likely to have occurred in many
or all of these systems. However, if migration occurred very early, prior to
planetesimal and embryo formation in the terrestrial planet region, then terrestrial planets could have
potentially formed after migration, with the giant planets in their current
configurations (e.g. \citealt{arm}). Our simulations focus on this scenario
(termed here ``in-situ formation''). If giant planet migration occurred after
planetesimals and embryos have formed, then our in-situ assumption does not
apply and there is likely to be radial migration of planetary embryos, driven by
giant planet migration (eg. \citealt{br3,raymond:2006,avi}). However, as there is currently
no clear consensus as to the most common timing of planetary migration, and no
evidence for the specific systems that we propose to study, each of our
simulations begin with the gas giants already fully formed and located in their
current positions. Simulations incorporating giant planet migration will be the
focus of the second paper in this series.

\subsection{Combining Dynamics and Chemistry}
The dynamical and chemical simulations were combined together as outlined in
\cite{bond:2009} whereby we assigned each embryo a composition based on its
formation location (hence pressure and temperature) and assumed that it
then contributed that same composition to the final terrestrial planet. The bulk
compositions of the final planets are simply the sum of each object they
accrete. Phase changes and outgassing were neglected and all collisions were
assumed to result in perfect accretion (i.e. no mass loss occurred).

\subsection{Stellar Pollution}
The issue of stellar pollution produced by terrestrial planet formation is of
great interest in extrasolar planetary systems. Pollution of the stellar
photosphere via accretion of a large amount of solid mass during planet
formation and migration has been suggested as a possible explanation for the
observed metallicity trend for known host stars \citep{la,g6,murray}. Thus we
determined the amount of material accreted by the host star during the current
terrestrial planet simulations and also determined the resulting change in
spectroscopic photospheric abundance. As in \cite{bond:2009}, any solid material
migrating to within 0.1AU from the host star is assumed to have accreted onto
the stellar photosphere. This material is then assumed to have been uniformly
mixed throughout the stellar photosphere and convective zone. Decreasing
convective zone mass with time, granulation within the photosphere and
gravitational settling and turbulence within the convective zone are again
neglected, resulting in the values determined here being the maximum expected
enrichments.

The mass of each element accreted onto the star was determined in the same way as for terrestrial planets, by summing the contributions of the individual embryos it accretes. As a reminder, the resulting photospheric elemental
abundance is given by:
\begin{equation}
{\rm [X/H]} = \rm log\left[\frac{f\rm_{X}}{f_{X, \bigodot}}\right]
\end{equation}\\

\noindent where [X/H] is the resulting abundance of element X after accretion of
terrestrial planet material, f$_{\rm X}$ is the mass abundance of element X in
the stellar photosphere after accretion and f$_{\rm X, \bigodot}$ is the mass
abundance of element X in the Solar photosphere (from \citet{murray}). Note that [X/H]
is still dependant on f$_{\rm X, \bigodot}$ as by definition it is taken
relative to the Solar abundance. Explicitly, f$_{\rm X}$ is given by:
\begin{equation}
{\rm f\rm_{X}} = \rm \left[\frac{m\rm_{X}}{m_{total}}\right]
\end{equation}\\

\noindent where m$_{\rm X}$ is the mass of element X accreted during the simulation, m$_{\rm total}$ is the total mass of the convective zone. Since the stellar composition is dominated by H, we can make the assumption that m$_{\rm total}$ $\sim$ m$_{\rm H}$. Thus:
\begin{equation}
{\rm f\rm_{X}} = \rm \left[\frac{m\rm_{X}}{m_{total}}\right] \sim \rm \left[\frac{m\rm_{X}}{m_{H}}\right]
\end{equation}\\

\noindent The same holds true for the solar values.

f$_{\rm X}$ values for the extrasolar planetary host stars were obtained via the
stellar abundances of \cite{gilli,be,ec,oxygen}, as these papers represent a
comprehensive, internally consistent catalogue of photospheric abundances for a
large number of known planetary host stars. The mass of the convective zone of a
star is known to vary with its mass, effective temperature (T$_{eff}$) and, to
some extent, its metallicity. Values for the masses of the convective zone for
each of the target stars were thus obtained from \cite{pin} using the
T$_{eff}$ values from \cite{ism04}. The convective zone masses are shown in
Table \ref{MCZ}. f$_{\rm X, \bigodot}$ values were obtained by utilizing the
solar abundances of \citet{asp} and a current solar convective zone mass of
0.03M$_{\bigodot}$ \citep{murray}.

\section{Results}
\subsection{Dynamical}
As the current simulations are only preliminary and are designed solely to
illustrate potential compositional differences within the final terrestrial
planets, a detailed examination of the simulation results is of minimal benefit.
However, terrestrial planets were found to form in \emph{all} simulations. 22 of
the 40 simulations produced two or more terrestrial planets within one system.
The general architecture of the resulting systems is shown in Figures
\ref{results1} - \ref{results5}.  In several cases, the orbital elements of the giant planets change slightly from the values shown in Fig. \ref{all_sys} due to the ejection of embryos from the system, and mutual perturbations in systems with multiple giant planets.

Several of these planets (e.g. namely those in the simulation for
HD4203) are simply embryos that have survived for the duration of the
simulation and have not accreted any additional material, but
essentially all others are grown from collisions among multiple embryos.

Of the ten planetary systems examined, only one (HD72659) is found to produce terrestrial planets with a median mass comparable to Earth (1.03 M$_{\bigoplus}$) with six of the 11 planets produced having masses equal to or greater than 1 M$_{\bigoplus}$. All other systems have median planetary masses less than 1 M$_{\bigoplus}$, and, excluding those of HD72659, only four out of 51 terrestrial planets in our simulations have masses equal to or greater than 1 M$_{\bigoplus}$.

With the exception of 55Cancri, terrestrial planets that form in our
simulations are located inwards of 1 AU. This is primarily a selection effect as
we are currently only simulating terrestrial planet formation interior to the
known giant planets, and the majority of systems we study here contain giant planets
orbiting within 2AU of their host star. 55Cancri is unusual in that its
outermost giant planet has a periapse larger than 5AU, a significant increase
over the other systems selected for study. That in turn dictated that the
embryos in the 55Cancri simulations initially be located between 1 AU and 5AU,
hence the larger semi-major axes of the planets that form there.

The terrestrial planets in these systems generally accrete the vast majority of their mass from their immediately surrounding area without a large amount of radial mixing occurring.  As such, the terrestrial planets in the simulations are expected to have compositions reflecting any radial trends within the disk. This amount of radial mixing, however, is likely to increase somewhat in future simulations that include a planetesimal swarm, and may greatly increase in simulations that include the effects of migrating giant planets.

The above results are not a definitive determination of the likely terrestrial planet population in the systems we are studying, as we are currently only considering late stage, in-situ terrestrial planet formation. That is, we are only considering formation that has occurred $\emph{after}$ the known giant planets have formed and migrated to their current orbits. As previously discussed, this is a valid approach as there is no consensus on the
timing or extent of migration within these systems and it provides us with a reasonable starting point to consider the chemical compositions
of possible terrestrial planets in those systems. However, giant planet migration can have a strong influence on terrestrial planet formation, in a worst-case scenario preventing terrestrial planets from forming, but it may also accelerate terrestrial planet formation and lead to significant radial mixing of material within the system, especially in systems with close-in giant planets \citep{raymond:2006,avi}. Simulations addressing the formation and chemical/dynamical evolution of terrestrial planets under the influence of giant planet migration are currently running and will be the focus of a future paper.

\subsection{Chemical\label{chem}}
The condensation sequences and abundances of solid species (normalized to the
abundance of the least abundant species) for three representative systems are shown in Figure \ref{combchem}. Plots for all other systems in order of increasing C/O value are available online only (Figures \ref{hscsumm1} - \ref{hscsumm3}). The 50\% condensation temperatures (i.e. temperature at
which half of the total amount of a given element has condensed) for
each of the systems studied is shown in Table \ref{EGP_Tc}. The final elemental abundances for all simulated planets and for
times studied are shown in Table \ref{all_summ} (available in full online).

Two very distinct types of condensation sequence are
produced for the systems studied here - those resembling the Solar condensation
sequence (HD27442, HD72659 and HD213240) and those in which carbide phases are
present within the disk (55Cnc, Gl777, HD4203, HD17051, HD19994, HD108874 and
HD177830). The C enrichment can further be classified as being low
(0.78$<$C/O$<$1.0), in which C and carbide phases are present within a
spatially narrow region of the disk at temperatures below $\sim$1800 K (55Cnc,
Gl777, HD17051 and HD177830), and high (C/O$>$1.0) where a broader
carbide-dominated region is stable for temperatures below $\sim$2300 K
and thus extends into the innermost reaches of the disk (HD4203, HD19994 and
HD108874). The implications of these variations in the distribution of solid
material are discussed in Section \ref{mass-dist}. The compositional
differences between these different classes of systems result in significantly
different compositions of the terrestrial planets that form in those systems,
and the characteristics of those planets will be discussed in the following
sections. Unless otherwise stated, all compositions shown are produced by disk
conditions at t = 5$\times$10$^{5}$ years. As previously stated, compositions
produced for disk conditions at this time were found by \cite{bond:2009} to
produce the best fit to observed Solar System compositions (based on agreeing with the compositions of Venus, Earth and Mars). Compositional
changes resulting from disk conditions at different times will be
discussed in Section \ref{EGPvary}.

\subsubsection{Earth-like Planets}
Before we can discuss the results produced by this work, we first need to define in detail what an "Earth-like" planet is. For the purposes of this study, it is taken to be at the most basic level a terrestrial planet whose composition is dominated by O, Fe, Mg and Si, with small amounts of Ca and Al and very little Carbon. Essentially, this refers to a terrestrial planet composed of Mg silicates and metallic Fe with other species present in relatively minor amounts. Water or other hydrous species may or may not be present. At a more detailed level, we have allowed for deviations of $\pm$25\% from the elemental abundances for the Earth listed in \cite{kandl} for the major elements (O, Fe, Mg, Si). Larger abundances were permitted in the minor elements (with the exception of C) due to their lower relative abundance within the terrestrial planets. Given the variations permitted on the individual abundances, a larger range of variations of 35\% was permitted in the geochemical ratios considered here (namely Mg/Si, Al/Si and Ca/Si). It is important to note that these accepted compositional variations span the full range of terrestrial planet compositions observed within the Solar System (i.e. under this taxonomy, both Venus and Mars would be classified based on their elemental composition as being Earth-like). As such, the definition of Earth-like should not be taken to mean identical to Earth in composition. Rather the classification refers to an elemental composition merely similar to that of the Earth.

Bearing that definition in mind, three systems (HD27442, HD72659 and HD213240)
were found to produce condensation sequences (and thus also terrestrial planets)
comparable to those of the Solar System. See Figure \ref{combchem} for the condensation sequence for HD27442. Schematic
representations of the abundances (for disk conditions at 5$\times$10$^{5}$
years) are shown in Figure \ref{earthlike} while the final elemental abundances for all times studied are shown in
Table \ref{all_summ}. From these it can be seen that for
HD27442, HD72659 and HD213240 the outermost terrestrial planets produced are grossly
similar in composition to known terrestrial planets. Their compositions are
dominated by O, Fe, Mg and Si with varying amounts of other elements. Upon
closer examination, however, large and important differences emerge, primarily
due to variations in the compositions of the host star and thus the initial
system itself.

Pronounced radial compositional variations can be seen in the simulated
terrestrial planets of all three systems. Planets located within $\sim$0.5AU for
all three systems contain large amounts of Al, Ca and
O, indicating that these planets formed from bodies containing the
high-temperature Al and Ca condensates (such as spinel and gehlenite) (see
Figure \ref{earthlike}). However, beyond $\sim$0.5AU, the refractory composition steadily decreases, producing
planets with compositions more closely correlated with that of the Earth,
dominated by O, Fe, Mg and Si, in the outer regions of the system. Planets produced within this transition zone between the two planetary types, however, are best described as being refractory-rich, Fe-poor silicate planets.

As expected, this difference is reflected in the planetary geochemical ratios,
as the planets located within the inner region have Mg/Si, Al/Si and Ca/Si
ratios well above the Solar System terrestrial planet values. However,
for planets located beyond the compositional transitional point, this is not the case. In this region, planetary Mg/Si, Al/Si and Ca/Si values are comparable to Earth
values. A steady transition
in composition between these two regions is seen for all three
systems (see Figure \ref{EGPratios}). This trend lies well above the observed
Earth fractionation line and is a result of the condensation of
refractory species in the innermost region and Mg-silicate species further out,
with relatively little radial mixing between the two regions during the
formation process. It is worth noting that while the Mg/Si values for the planets produced in the system of HD213240 are still just within the upper limits of the accepted Earth-like Mg/Si values, the higher modeled planetary values are a result of the increased Mg/Si value
of HD213240 itself (Mg/Si$_{\bigodot}$ = 1.05, Mg/Si$_{\rm Gl777}$ = 1.48). This results in a system with olivine as the major Mg silicate condensate and little pyroxene present.

The model terrestrial planets in the three systems with condensation sequences comparable to that of the Solar System (HD27442, HD72659 and HD213240) can thus be characterized as being essentially similar in composition to Ca- and Al-rich inclusions (CAI's) (for the innermost planets) and Earth-like (for the outermost planets).

\subsubsection{C-rich Planets} In systems with C/O values close to or
above 0.8, the planets that form can begin to incorporate carbon as a
significant planet-building material, in the form of graphite, SiC and TiC,
which is a significant difference compared to what occurs in the Solar System.
Seven such systems were selected for the current study: four with low
carbon enrichment (0.8$\lesssim$C/O$<$1: 55Cnc, Gl777, HD17051 and HD177830) and
three with a significantly higher C content (C/O$>$1: HD4203, HD19994 and
HD108874). The final elemental abundances for all times studied are shown in
Table \ref{all_summ}.

\subsubsection{Low C-enriched Planets}
\emph{Gl777:} For disk conditions at t = 5$\times$10$^{5}$ years, Gl777 produces
Earth-like terrestrial planets with minor C enrichment. See Figure \ref{combchem} for the condensation sequence for Gl777 and Figure \ref{lowC} for a schematic of the resulting planetary elemental abundances. The final elemental
abundances of the refractory lithophile and siderophile elements are similar to those of Earth with the simulated planets displaying a marginal
enrichment in Mg ($\sim$3wt\%) and a depletion in Si ($\sim$1wt\%) and
Fe ($\sim$2wt\%) compared to Earth values. The most volatile species (P,
Na and S) are enriched over Earth, which is likely a result of the fact that we
do not consider volatile loss during accretion, which is expected to be
significant as for the Solar System simulations.

The geochemical ratios of Mg/Si and Al/Si are enriched compared to those of the
Earth, although they are still in accepted range for an Earth-like planet in this study (see Figure \ref{Gl777ratios}). As for HD213240, this
increase in the planetary Mg/Si values over that observed for the Earth is due to the slight increase the Mg/Si value
of Gl777 itself (Mg/Si$_{\bigodot}$ = 1.05, Mg/Si$_{\rm Gl777}$ = 1.29). In
turn, this produces a system containing nearly equal amounts of olivine and
pyroxene (compared to the pyroxene dominated Solar disk) and results in
Mg-enriched planets. Note that we do not see such a large spread in planetary Al/Si values as was observed in Figure \ref{EGPratios} as Gl777 does not contain any terrestrial planets in the region dominated by Al-rich CAI-like material (i.e. within $\sim$0.5 from the host star).

The Ca/Si values, however, are lower than those of Earth (Ca/Si$_{Gl777}$ =
0.07, Ca/Si$_{\bigoplus}$ = 0.11), although once again falling within our accepted range of values. This variation is primarily due the fact that there is relatively less Ca within the system. Ca is one of the least enriched
elements within Gl777 ([Ca/H] = 0.10 vs. [Al/H] = 0.34), resulting in a relative
Ca depletion within
the solid material. The variation in the abundances of the host star is
reflected in the lower Ca/Si value of the final planets produced. This
difference is certainly no larger than that observed for the Solar System
simulations previously discussed \citep{bond:2009}.

Although the C/O ratio for Gl777 is slightly below 0.8
(C/O$_{Gl777}$=0.78), a carbon-rich region is still predicted to occur at lower
pressures (10$^{-5}$ bar and below), resulting in the production of a narrow
carbide-dominated region (extending from 0.77 to 1.13AU at 5$\times$10$^{5}$
years). As a result, the terrestrial planets produced in Gl777 are all
slightly enriched in C compared to the Earth (containing up to
0.5 wt\% C). This enrichment decreases for disk conditions at later times as the
C region migrates inwards, interior to the feeding zones of the terrestrial
planets.

Given these simulated compositions, the terrestrial planets that could form around Gl777 are best characterized as being Earth-like with a minor C-enrichment in their chemical
composition. It is also interesting to note Gl777 is very close to the average extrasolar planetary host star values of Mg/Si and C/O (1.29 and 0.78 respectively, compared to the average values of 1.32 and 0.77). This result implies that terrestrial planets in an ``average'' extrasolar planetary system would have compositions comparable to that of our own Solar System but moderately enriched in Mg, and with a potentially minor C-enrichment.

\emph{55Cnc:} Like Gl777, 55Cancri produced terrestrial planets that are C-enriched, containing up to 9.28 wt\% C (see Figure \ref{lowC}). They
are similar to Earth in that they are expected to be dominated by Mg-silicate
species, with metallic Fe also present. However, the planets of 55Cnc are also enriched in both S and Mg beyond our acceptable limits to be called an Earth-like planet. This results in modeled planets with Mg/Si values above that of the Earth and Ca/Si values well below. As for HD213240 ad Gl777, this high
planetary Mg abundance is caused by the fact that 55Cancri is highly enriched in
Mg ([Mg/H] = 0.48), resulting in olivine becoming the major silicate species
present within the disk and thus producing the high Mg/Si value observed.

Although the disk of 55Cancri contains a C-rich zone, producing C-enriched terrestrial planets for disk conditions at t=5$\times$10$^{5}$, for disk conditions at later times, none of the simulated planets are predicted to contain any C because the C-rich region moves interior to the terrestrial planet zone in that system.  All of the terrestrial planets in 55Cnc are located between 1.5 and 4AU while the C-rich zone is located between $\sim$1375 and 713K, corresponding to a radial distance of 0.46 and 1.48AU (for disk conditions in the \cite{hersant} model at 5$\times$10$^{5}$ years). Thus the primary feeding zones for each of the planets are located on the outer edge of the C zone at 5$\times$10$^{5}$ years, and well beyond it at later times. Thus for disk conditions at later times, planets best described as Mg-rich Earths are produced. The location of the C zone also implies that the four inner known giant planets of the 55Cnc system (located at 0.038AU, 0.115AU, 0.24AU and 0.781AU) should contain significant amounts of C, both in their solid cores and in their atmospheres \emph{if} they formed at or in close proximity to their current orbital locations and also depending on the exact time of their formation. Given the variation in the location of the C rich zone with time, it is expected that terrestrial planets containing some C would also be produced in the current simulations for disk conditions at later times if a time varying equilibrium composition for the solid material was incorporated into the simulations rather than the ``snapshot'' approach taken here.

\emph{HD17051:} Although the disk of HD17051 does contain a C-rich zone, for the disk conditions at t = 5$\times$10$^{5}$ years the simulated
terrestrial planets do not contain any C, instead resembling the planets of Solar System-like systems previously discussed (see Figure \ref{lowC}). The innermost
planets (within $\sim$0.5 AU from the host star) are dominated by Al and Ca, resembling CAI's, while the outer planets are more Earth-like,
consisting of Mg-silicates and metallic Fe. They are still significantly enriched in Fe, mostly due to their primary feeding zone location (and subsequent composition) and the high metallicity of HD17051 itself ([Fe/H] = 0.26).

However, \textbf{for planetesimals initially forming under }disk conditions at later times the planetary
composition for all simulated planets changes to more closely resemble a C-enriched Earth-like planet, with planets dominated by O, Fe, Mg and Si and a
significant amount of C.
Up to 4.37 wt\% C is predicted to exist in the planets for the disk conditions
at 3$\times$10$^{6}$ years. These planets are essentially C-enriched Earths,
containing the same major elements in geochemical ratios within limits to be considered Earth-like, but
also an enhanced inventory of C, primarily accreted as solid graphite. As for
55Cnc, it is expected that if were we to incorporate time-varying equilibrium compositions into our models that we would see C occurring in the
terrestrial planets for all simulation times.

\emph{HD177830:} HD177830 has the highest Mg/Si (and Al/Si) ratio of any system
simulated. This enrichment alters the compositions of major silicate species
present within the disk. While the Solar System should have condensed both
olivine and pyroxene between 0.35 and 2.5 AU, HD177830 is dominated by olivine
beyond 0.3 AU and contains only a small region where pyroxene is predicted to
coexist. This unusual composition is reflected in the final planetary abundances
as the planets contain large portions of Mg (up to 22.33 wt\%) (see Figure \ref{lowC}) and have a mean
Mg/Si value of 1.71, well above Earth values (Mg/Si$_{\bigoplus}$ = 1.01).

Al is also similarly enriched (up to 31.74 wt\%), again because of the high Al
abundance of the host star and thus the initial system itself. Other refractory
and lithophile elemental abundances within the final planets are comparable to
that of the Solar System. Of the 10 planets produced in our
simulations, five also contain trace amounts of C (up to 2.96 wt\%). This
increases for disk conditions at later times as the C-rich region migrates
inwards from its initial location at 0.63 - 1.66 AU (at t = 5$\times$10$^{5}$
years), through the primary feeding zone, producing planets with increased
C-abundances (up to 9.8 wt\%). The planets of
HD177830 can best be described as being Mg- and Al-rich silicate planets with some
C-enrichment. Such a Mg dominated planetary composition would undoubtedly alter
the interior structure and processes of the planets themselves. Such
considerations will be discussed further in Section \ref{implications}.

\subsubsection{High C-enriched Planets}\emph{HD19994, HD108874 \& HD4203:} All three systems all have C/O
values above 1.0 (1.26, 1.35 and 1.86 respectively). In all three systems, the
inner regions of the disk are completely dominated by refractory species
composed of C, SiC and TiC, as opposed to the Ca and Al-rich inclusions
characteristic of the earliest solids within our Solar System (see Figure \ref{combchem} for the condensation sequence for HD4203). Significant
amounts of metallic Fe are also present within these systems. As all three
systems produced terrestrial planets located within 0.7AU from their host star,
these unusual inner disk compositions produced terrestrial planets primarily
composed of C, Si and Fe. HD19994 produced terrestrial planets composed almost
entirely of SiC and metallic Fe and containing up to 37 wt\% Si and between 31
and 63 wt\% C, over 100 times more C than is estimated for Earth (see Figure
\ref{highC}, upper left). The outermost terrestrial planet for HD19994 does
contain significant amounts of O and Mg, primarily because its feeding
zone, although still undoubtedly dominated by C, is also rich in pyroxene. This
presence of a Mg silicate species produces a slightly more varied composition
for a single simulated planet in one of the four simulations completed.

More extreme deviations occur when we consider the planets formed for HD108874
and HD4203. Both of these systems have considerably wider graphite dominated
regions, extending from 1.5AU to within 0.1AU (for disk conditions at
5$\times$10$^{5}$ years). As a result, terrestrial planets are found to be
composed almost entirely of C, Si and Fe (see Figure
\ref{highC}). Both HD108874 and HD4203 produced
terrestrial planets composed almost entirely of C and SiC and containing between 67
and 74 wt\% C (for midplane conditions at 5$\times$10$^{5}$ years). Only the outermost planet of HD108874 contained
less than 50 wt\% C (30.54 wt\% C), with the remainder of the planet consisting
of Fe and Si. For disk conditions at later times, Mg and O are also present in all planets produced
for both systems, again due to the incorporation of pyroxene and olivine into
the planetary feeding zones. It should be noted that most of the terrestrial
planets formed in the HD4203 simulations are single embryos that survived for
the duration of the simulation but did not accrete any other solid material.
Terrestrial planets within these systems are unlikely to have compositions
resembling that of any body we have previously observed. The possible
implications of these types of planetary compositions will be discussed in
Section \ref{implications}.

\subsection{Stellar Pollution}
The average change in stellar photospheric abundances produced by accretion
of solids onto the star for disk conditions at 5$\times$10$^{5}$ years
is shown in Table \ref{EGPpollution} for each system studied.
The majority of systems experienced minimal increases in photospheric abundances
as a result of accretion during terrestrial planet formation. The largest
elemental enrichments occurred for the most refractory elements (Al, Ca, Ti, Ni
and Cr), primarily because refractory material is initially
located closest to the star itself. The simulations for HD72659 produced many of
the most significant enrichments for all elements examined. This is primarily
due to the large amount of mass accreted (2.640 M$_{\bigoplus}$) and the lower
estimated mass of the stellar convective zone (0.0112M$_{\bigodot}$). However,
as previously discussed, mixing within the stellar radiative zone is
not incorporated into the current approach. As such, these values are
upper limits for those stars with low mass convective zones and large radiative
zones. Furthermore, HD72659 (along with Gl777) has the highest degrees
of radial mixing and subsequently also accreted the largest amount of solid
material onto their host stars. This resulted in a larger change in the
stellar photospheric abundances than for other systems.

All of the predicted abundance changes are below the errors of current
spectroscopic surveys (e.g. $\pm$0.03 for \cite{fv}), meaning that definitively
observable elemental enrichments are not necessarily predicted by our
terrestrial planet formation simulations. Of course, inclusion of a
swarm of planetesimals within the formation simulations and migration of the
giant planets is expected to increase the amount of material accreted by the
host star and thus also the predicted stellar abundances. However, these
increases are expected to be minor and would thus still
result in only marginal increases in the observed elemental abundances

\section{Implications and Discussion\label{implications}}

\subsection{Mass Distribution\label{mass-dist}}
Radial midplane mass distributions based on the equilibrium condensation
sequence were calculated for each system. As composition is correlated to a
specific radial distance within the midplane (via the \cite{hersant} model), the
total mass of solid material present interior to a given radial
distance within the disk is given by summing the masses contained within
each annulus of the disk for which composition is calculated:

\begin{equation}
{\rm Mass\:of\:solid\:material} = \rm {\Sigma_{i=0.1AU}^{i=5.0AU}}2{\pi}r_{i}drM_{solid,\:i}
\end{equation}\\

\noindent where M$_{\rm solid,\:i}$ is the mass of solid material determined by the chemical model to be located in an annulus of width dr at r$_{\rm i}$. The mass of solid material possible is determined by using the minimum mass solar nebula with a gas surface mass density profile varying as r$^{-3/2}$, normalized to 1700 gcm$^{-2}$ at 1 AU. Note that this approach only considers equilibrium driven condensation and neglects other processes that may migrate material and alter the mass distribution.

Based on this calculation, the most carbon-rich systems simulated have
significant differences in their mass distributions compared to
other systems. The combination of a broad zone of refractory carbon-bearing
solids in the inner regions and the relatively small amount of water
ice that condenses in the outer regions of these systems suggests that
C-rich systems have significantly more solid mass located in the inner regions of the disk
than for a Solar-composition disk. This can be seen in Figure \ref{massdist} which shows the distribution of
solid mass within each system normalized to a solar composition disk for disk
conditions at t = 5$\times$10$^{5}$ years.

From Figure \ref{massdist}, the system with the highest C/O value (HD4203) clearly contains significantly more mass in the innermost regions of the disk than a disk of solar composition does. However, the total mass of solid material produced by the current approximations is only 18.4 M$_{\bigoplus}$. As such, it is not clear that a giant planet core composed of refractory C-rich species could be produced within several AU of the host star, allowing for giant planet formation to occur much closer to the host star than previously thought. If significantly more mass were present within the disk than is currently estimated, such a scenario may be possible and would obviously alter the extent and nature of planetary migration required within these systems as it would no longer be required that a planet located at 1-2AU originally formed at 5AU and migrated inwards. Alternatively, if insufficient mass is available for Jovian core formation, production of large terrestrial planets in this region may proceed faster and with greater ease, thus increasing the chance of forming detectable terrestrial planets.

It can be seen from Figure \ref{massdist} that the planetary system with the most Solar-like composition (HD 72659) has a mass distribution similar to that of the Solar System. The enrichment observed over the solar mass distribution within $\sim$0.5AU from the host star is due to the higher Mg/Si value for HD72659 resulting in the condensation of more Mg silicates. This enhancement is more obvious for the system with the highest Mg/Si value (HD177830). Likewise, the refractory rich system HD27742 also contains more mass in the inner regions of the disk than for a Solar abundance. This is due to the high abundances of several refractory elements ([Al/H]=0.53, [Na/H]=0.41, [Ca/H]=0.12). The variation in mass observed for all systems at $\sim$4.8AU is due to the condensation of hydrous species in various amounts relative to the solar composition disk.

These results imply that the use of an alternative initial mass distribution may be required for planetary systems with high C/O values. Although the planetary formation simulations presented here utilized the classical Solar-System based mass distribution, such an approach is still valid for the illustrative purposes of the current study. The full implications of these results need to be examined in future work by using alternative mass distributions for extrasolar planetary formation simulations for both gas giant planets and smaller terrestrial planets.

\subsection{Timing of Formation\label{EGPvary}}
Specific planetary compositions have been found to be highly dependent on the
time selected for the disk conditions, especially for those systems containing
C-dominated regions. This is primarily due to the low degree of radial mixing
encountered within the simulations. As a result, as conditions within the area
immediately surrounding the planet evolve, the composition of the solid
material and thus the final planet itself drastically change. For disk
conditions at later times,the simulated planetary compositions evolve to more closely resemble those of
the Solar System. They become dominated by Mg silicate species and metallic Fe.
Terrestrial planets in Solar-like systems attain more hydrous material. Variations in composition with disk condition times are most noteworthy for
those planets dominated by refractory compositions (such as the inner planets of
HD72659 and HD27442). Under later disk conditions, these planets experience a
complete shift in their composition, losing the majority of their refractory
inventory to be composed primarily of Mg-silicates (olivine and pyroxene).
Therefore if solid condensation and initial planetesimal formation occurred significantly
later, we would expect to observe predominantly Mg-silicate and metallic Fe
planets with enrichments in other elements (such as C) depending on the exact
composition of the system. Although disk conditions at 5$\times$10$^{5}$ years
provide the ``best fit'' for Solar System simulations (based on fitting the compositions of Venus, Earth and Mars) and are thus utilized
here, it remains to be seen whether or not disk conditions at this time provide
an accurate description of the conditions under which planetesimals and
embryos formed in other planetary systems. Therefore, we require a more
detailed understanding of the timing of condensation and planetesimal and embryo
formation within protoplanetary disks to be able to further constrain the
predicted elemental abundances.

Similarly, as the disk evolves, the various condensation fronts migrate closer to the host star. For example, the water ice line for Gl777 migrates from 7.29 AU for midplane conditions at 2.5$\times$10$^{5}$ years to 1.48AU for midplane conditions at 3$\times$10$^{6}$ years. Similar degrees of migration also occur for the condensation fronts of other species (such as the Mg silicates olivine and pyroxene). In effect, the change in location of the condensation fronts alters the mass distribution within the disk, increasing the mass present in both the very closest regions of the disk ($<$ 1AU) as the refractory species are replaced by the more abundant Mg silicates and in the outer most regions ($>$ 3AU) as water ices appear. The full effects of this change will require formation simulations to be run with alternative mass distributions but it is thought that such conditions will increase the efficiency of forming close-in terrestrial planets and/or the mass of the resulting planets. Additionally, it will also allow for efficient terrestrial planet growth in the outer regions, possibly to the extent of forming gas giant cores. Given that the average disk lifetime is $\sim$3Myr, not all disks will reach the final chemical compositions modeled here. Thus it remains to be seen not only whether or not sufficient solid mass would be retained during the evolutionary process for Jovian cores to develop but also whether or not core formation can occur before the disk is cleared out.

\subsection{Detection of Terrestrial Planets} The results of this study are of great importance for the design of terrestrial planet finding surveys. Our simulations, while preliminary, suggest that terrestrial planets can be stable in a wide range of extrasolar planetary systems. Four distinct classes of planetary composition have been produced by the current simulations: Earth-like, Mg-rich Earth-like, refractory (compositions similar to CAI's) and C-rich. These planetary types are primarily a result of the compositional variations of the host stars and thus the system as a whole. Based on their observed photospheric elemental abundances, the majority of known extrasolar planetary systems are expected to produce terrestrial planets with compositions similar to those within our own Solar System. Therefore, systems with elemental abundances and ratios similar to these (e.g. HD72659) are ideal places to focus future ``Earth-like'' planet
searches.

Based on our dynamical simulations, the masses of the terrestrial planets produced are too low to be detected by current radial velocity surveys. However, many of the simulated planets are in orbits (assuming the simulated inclinations are correct) that would place them within the prime target space for detection by the Kepler mission. Designed to detect extrasolar planets via transit studies, Kepler is the first mission that has the potential of detecting Earth-mass (and lower) extrasolar planets located within the habitable zone of a planetary system. It has the sensitivity to detect the transit of an Earth mass body within 2AU from the host star and a Mars mass body (0.1Me) within 0.4AU. Single transit events may not be sufficient to positively identify the presence of a planet, although planets at ~1 AU should transit ~3-4 times during the lifetime of the kepler mission, thus reducing the likelihood of contamination in the data. The vast majority of the terrestrial planets formed here (with the exception of the lowest mass, highest semimajor axis planets) are well within this range and thus may be detectable if they are indeed present within these systems (assuming the system is aligned such that transits can be observed from orbit). Only HD4203 produces no potentially detectable planets, based on their predicted masses. Thus it is likely that we will have an independent check of extrasolar terrestrial planet formation simulations within the next 5 years (but not necessarily for these systems, only for a range of systems which may be similar to these). Such information will be vital for further refinement of planetary formation models for both giant and terrestrial planets. Obtaining compositional information about such terrestrial planets, however, will be more difficult as the size and location of the predicted planets will prohibit direct spectroscopic studies. It is also unlikely that the terrestrial planets will contain atmospheres large enough to be detected and characterized by transit surveys. As such, detailed extrasolar terrestrial planetary chemical compositions will remain unknown for the foreseeable future.

In addition to detection via transit surveys, attempts are also being made to
obtain direct images of extrasolar planetary systems. One such example is
Darwin, a proposed ESA space based mission that would utilize nulling
interferometry in the infrared to directly search for terrestrial extrasolar
planets. The compositional variations outlined here are likely to influence our
ability to successfully detect these planets. Carbon-rich asteroids are known to be highly non-reflective. For example, 624 Hektor
(D-type asteroid) has a geometric albedo of 0.025 while 10 Hygiea (C-type
asteroid) has a geometric albedo of 0.0717 (compared to a geometric albedo of 0.367 for the Earth and 0.113 for the moon). As both of these asteroids are assumed to be
carbon-rich, it
is likely that the carbon-rich planets identified here are similarly dark. Thus
it is expected that searches for these planets in the visible spectrum will be
difficult due to the small amount of light reflected by these bodies. However, a
lower albedo results in greater thermal emission from a body, suggesting that
the infrared signature from these planets would extend to shorter wavelengths than
corresponding silicate planets. As a result, infrared searches (such as that of Darwin) are
ideally suited to detect carbon-rich terrestrial planets and thus should be
focused on stellar systems with compositions similar to that of the C-rich stars
identified here to maximize results. Of course, this conclusion neglects any possible effects of a planetary atmosphere.

\subsection{Hydrous Species}
As one would intuitively expect, hydrous material (water and serpentine in the
current simulations) is primarily located in the outer, colder regions of the
disks. This corresponds to beyond $\sim$7.3AU for disk conditions at
2.5$\times$10$^{5}$ years and beyond
$\sim$1.4AU for disk conditions at 3$\times$10$^{6}$ years for all compositions examined. As a result of this
distribution, terrestrial planets forming in the inner regions of a given
planetary system are unlikely to directly accrete significant amounts of
hydrated material. In the current in-situ simulations, none of the simulated
terrestrial planets directly accrete any hydrous species for disk conditions at
5$\times$10$^{5}$ years. As the composition of the planetesimals is dictated by the thermodynamic conditions of the disk at the time of condensation only, if any of the simulated planetesimals were to condense/form at later times, they would be more likely to be
water-rich given that the `snow line'\footnote{Note that here `snow line' refers to the location within the disk where the thermodynamic conditions are such that water ice condenses (i.e. where T = 150K).} migrates inwards as the disk cools,
producing a greater overlap between their feeding zones and the water-rich
region of the disk.

However, a far greater effect is expected to be produced by migration of giant planets within the system \citep{raymond:2006,avi}. Migration of
a giant planet has been shown to be capable of driving a large amount of material from the outer regions of a disk inwards. In the case of
hydrous material, this has been found to result in water-rich terrestrial planets being formed both interior and exterior to the giant planet
\citep{raymond:2006,avi}. The full extent of this increase in water content will be examined by a suite of simulations incorporating giant migration that
are currently running and will be the focus of future work.

In the Solar System, it has been hypothesized that the Earth's water could have originated from hydrated material in the region where the asteroid belt now lies, or from comets beyond the orbit of Jupiter.  While belts of debris resembling the asteroid belt could exist interior to the giant planets in several of the extrasolar systems and may be incorporated into the terrestrial planets, as noted above, none of that material is expected to be hydrated at early times in the disk. We can not address the issue of cometary delivery of water in these simulations, as bodies beyond the orbits of the giant planets are not presently included in our dynamical models.

Water can also potentially be incorporated into a planetary body via
adsorption onto solid grains within the disk \citep{drake:2005}. While this
process has not been considered in our current simulations, it is
possible that there will be some water delivered during the formation
process to the terrestrial planets produced in the Earth-like systems (HD27442,
HD72659 and HD213240) as the solid grains are bathed in water vapor over the
entire span of the disk. This same process will likely not be as effective at
delivering water to the C-rich systems (55Cnc, Gl777, HD4203, HD17051, HD19994,
HD108874 and HD177830) as they only have water vapor present at temperatures
below $\sim$800 K. This temperature range corresponds to beyond a radial
distance of $\sim$1.2AU for \cite{hersant} midplane conditions at
2.5$\times$10$^{5}$ years and $\sim$0.2AU for midplane conditions at
3$\times$10$^{6}$ years. As few terrestrial planets in our simulations
accrete material from beyond 1.2AU, it is expected that C-rich planets forming
early in the lifetime of the disk will remain dry without additional water being
incorporated into the planets via adsorption. Thus it appears that
terrestrial planets are likely to obtain some amount
of water (through giant planet migration mixing the disk, variations in
composition with time i.e. heterogeneous accretion, adsorption and
exogenous delivery), while those within Solar-like systems may receive more water and other hydrous species than terrestrial planets within C-rich systems.

\subsection{Planetary Interiors and Processes}
Given the wide variety of predicted planetary compositions, a similarly diverse
range of planetary interiors structures is also expected. To better
quantify this, we examined three specific cases: a 1.03M$_{\bigoplus}$
Earth-like planet located at 0.95AU (HD72659), a 1.22M$_{\bigoplus}$ Mg-rich
Earth-like planet located at 0.43AU (HD177830) and a 0.47M$_{\bigoplus}$ C-rich
planet located at 0.38AU (HD108874). Approximate interior structures for each
were calculated using equilibrium mineralogy for a global magma ocean with P =
20GPa and T = 2000$^{\circ}$C. Equilibrium compositions at these conditions have
been found to produce the best agreement between predicted and observed
siderophile abundances within the primitive upper mantle of the Earth
\citep{drake}. Elemental abundances were taken from the results discussed in
Section \ref{chem}. Resulting mineral assemblages were sorted by density to
define the compositional layers. Note that this assumes that a planet undergoes complete melting and differentiation. Approximate planetary radii were obtained from
\cite{sotin} based on planetary mass. These planetary radii are based on
silicate planetary equations of state and as such are unlikely to completely
describe the C-rich planets modeled here. However, no studies have
considered such assemblages, forcing us to assume a silicate based approximate
radius. Density variations at high pressures were not considered in defining the
depths of various layers. Although important for planetary evolution processes such as mantle stripping, the effects of large impacts (such as the moon forming impact) are
also neglected. The
resulting interior structures (shown to scale) can be seen in Figure
\ref{interior}.

The interior mineralogy and structure of one of our model planets
orbiting HD72659 is similar to Earth. It contains a pyroxene and feldspathic
dominated crust ($\sim$133 km deep) overlying an olivine mantle ($\sim$985 km
deep) with an Fe-Ni-S core (radius $\sim$ 4930 km). The crust is thicker than
seen on Earth as we are currently neglecting density and phase changes. Given
its structure and comparable mineralogy, we would expect to observe planetary
processes similar to those seen on Earth. The planets location within the
habitable zone of the host star suggests that a liquid water ocean is feasible, provided sufficient hydrous material can be delivered. Melting conditions and magma
compositions are expected to be comparable and it is feasible that a liquid core
would develop, resulting in the production of a magnetic dynamo. In general, based on their mass and
composition, the terrestrial planets of HD72659 are likely to have structures
and mineral assemblages similar to those observed in our system.

The simulated planet for HD177830 (the system with the highest Mg/Si value) is depleted in Si, relative to the Earth, resulting in high spinel and
olivine content in the mantle (resembling that of type I kimberlites) and a
thicker mellite and calcium dominated crust than found for HD72659
($\sim$309 km deep). The core would produce a considerable amount of heat
via potential energy release during differentiation, potentially
producing melts with compositions similar to komatiite (dominated by olivine
with trace amounts of pyroxene and plagioclase). Volcanic eruptions would be
comparable to basaltic flows observed on Earth due to the low silica content of
the melt. However, given the thickness of the crust, extrusive volcanism and
plate tectonics are unlikely to occur as high stress levels would be required to
fracture the crust. Producing and sustaining such stresses would be challenging.
Therefore, it is questionable whether or not a planet with this composition and
structure would be tectonically active for long periods of time. Given
the similar composition and size of the core compared to Earth, a
magnetic dynamo is still expected to be produced within the core.

Finally, carbide planets are expected to form around HD108874. The resulting composition and structure is unlike any known planet. Its small
size, refractory composition and possible lack of radioactive elements (due to the potential absence of phosphate species, common hosts for U
and Th, and possible lack of feldspar and carbonates, the common host of K) will inhibit long-term geologic activity due to the difficulty of melting the
mantle. Only large amounts of heat due to core formation and/or tidal heating would be able to provide the required mantle heating. Once all the
primordial heat has been removed, it is unlikely that the mantle would remain molten on geologic timescales. Until that time, given the buoyancy
of molten carbon, volcanic eruptions would be expected to be highly enriched in C. The core is also expected to be molten, thus making it likely
that a magnetic dynamo would be produced \citep{gaidos}. Note that this assumes that sufficient heat is initially available to melt the body and allow for differentiation and core formation to occur in the first place. In essence, although initially molten and probably active, old carbide planets of this type would be geologically dead.

Incomplete mixing of material accreted at later times is likely to result in deviations from the equilibrium picture presented here.
For example, accretion of oxidized and water-rich material late in the formation process may result in a stratified redox state and water-rich
crust as observed for the Earth. Unfortunately, it is not possible to determine these effects with current models as it requires a level of
understanding of the impact and accretion process (e.g. mantle mixing, fragmentation) on small planetary bodies that we currently do not have.

These results are also key for super Earth studies such as that of \cite{plate1} and \cite{plate2}. Previous simulations have assumed
Earth-based compositions and structures. Based on the present simulations, a wide variety of both are possible and will need to be considered.

\subsection{Planet Habitability\label{habit}}
The habitable zone of a planetary system is defined as being the range of
orbital radii for which water may be present on the surface of a planet. For the
stars considered here, that corresponds to radii from $\sim$0.7AU to
$\sim$1.45AU. The vast majority of the planets produced by the current
simulations orbit interior to this region (exterior in the case of 55Cnc) and
thus are unlikely to be habitable in the classical sense. 10 planets
are produced within the classical habitable zone, existing in orbits extending
from 0.70AU to 1.19AU. Seven on these planets are formed in Solar-like systems
(HD24772 and HD72659) and have compositions loosely comparable to that of Earth. As
such, we feel that these systems (and others similar to them) are the ideal
place to focus future astrobiological searches as they may not only contain planets
with compositions similar to that of Earth but also exist in the biologically favorable region of the
planetary system.

Of the seven C-rich systems, only two produced planets within the habitable
zone. Gl777 formed two terrestrial planets within the habitable zone while
HD19994 formed a terrestrial planet at 0.70AU, just at the inner edge of the
habitable zone. All other planets are located well inside the required radii. Both of the habitable planets around Gl777 are
C-enriched Earth-like planets, making them potential sites for the development
of life. The single habitable planet produced by the HD19994 simulations is
dominated by C, along with O, Fe, Si and Mg. It is unclear whether such a
composition would be favorable to life. Additionally, the low planet mass
(0.06M$_{\bigoplus}$) further makes it unlikely that this particular simulated
planet could ever actually host life. As such, under the current definition of
habitable, we conclude that of the seven C-rich systems currently simulated,
Gl777 has the best chance of supporting life, but this is by no means guaranteed.

\subsection{Biologically Important Elements}
In addition to water, complex life (as we know it) also requires several key elements to exist. The six essential elements are H, C, N, O, P and
S. As was the case for the Solar System simulations discussed in \cite{bond:2009} none of the planets accreted any N and are also lacking in H. The terrestrial planets formed in the Solar-like systems contained
various amounts of O and P but some were deficient in S and all were laking C, as for the Solar System simulations. The most C-rich systems, on the other hand, were
lacking in O, P and S.

Thus it is clear that for life as we know it to develop on any of the terrestrial planets formed in the current simulations, significant amounts of several
elements must be supplied from exogenous sources within the system. All elements may be supplied from the outer, cooler regions of the disk.
Thus it is possible that migration or the radial mixing of cometary-type material into the terrestrial planet region may produce planets with the necessary elements for life to develop.
As for the Solar System simulations, all biologically required elements would be introduced in a form that could potentially be utilized by early life. This
is especially intriguing for those planets located within the habitable zone. On the other hand, alternative pathways could potentially develop for the
formation of an alternative biologic cycle without requiring the same six elements.

\subsection{Host Star Enrichment}
As previously discussed, stellar photospheric pollution has been suggested as a possible explanation for the observed high metallicity of
extrasolar planetary host stars \citep{la,g6,murray}. The current simulations, though, do not support this hypothesis. Enrichments are produced
primarily in Al, Ca and Ti, not Fe as is required by the pollution theory. Furthermore, relatively small masses of solid material are accreted
by the host stars during planet formation, suggesting that insufficient material is accreted to produce the observed enrichments. Thus unless
migration of the giant planets can systematically result in accretion of giant planets by the host star, our results agree with with previous authors
(e.g \citealt{s1,s2,s04,s05,fv}) in finding that the observed host star enrichment is primordial in origin.

Our simulations also imply that enrichments due to stellar pollution are most likely to be observed for the refractory elements in high
mass stars with low convective zone masses. This suggests that surveys for pollution effects caused by terrestrial planet formation should focus
on Ti, Al and Ca abundances in A-type and high mass F-type stars as they are expected to have the lowest convective zone masses. However, more
detailed simulations of the fate of material accreted into radiative zones need to be undertaken to support this hypothesis.

\section{Summary}

Terrestrial planet formation simulations have been undertaken for ten
different extrasolar planetary systems. Terrestrial planets were found to
form in all systems studied, with half of the simulations producing
multiple terrestrial planets. The simulated planets are possibly detectable by Kepler (if the system orientation is favorable), thus potentially allowing for future independent verification of formation simulations.

The compositions of these planets are found to vary greatly, from those
comparable to Earth and CAIs to other planets highly enriched in carbide
phases. These compositional variations are produced by variations in
the elemental abundances of the host star and thus the system as a
whole, {with the Mg/Si and C/O ratios being the most important for
determining the final planetary compositions. Based on this, it is expected
that C-rich planets will comprise a sizeable portion of extrasolar terrestrial planets and need to be
considered in significantly more detail. These compositions are highly dependant
on the disk conditions selected for study, requiring us to develop a more
detailed understanding of the timing of planetary formation within these
systems.

Given the wide variety of compositions predicted, it is also likely that
planetary mineralogies and processes within these planets will be
different from those of our own Solar System. Compositions range from
planets dominated by Fe and Mg-silicate species to those composed almost
entirely of Fe and C. These compositional variations are likely to generate
differences in detectability with C-rich planets being easier to detect via
infrared surveys such as the Darwin mission due to their lower albedo
and hence larger IR emission.

The most habitable planets are expected to be those forming in systems with
compositions similar to Solar. The simulated compositions make these planets ideal targets for future
astrobiological surveys and studies. Planets in the C-rich systems that
we model are likely to be lacking water and generally located interior to the
habitable zone, making such planets unfavorable for the development of life as
we know it.

Finally, pollution of the host star by the planetary formation process appears
to be negligible for the majority of systems. Enrichments are produced only for
those stars with the least massive convective zones and even then only in the
most refractory elements (Ti, Al and Ca). Therefore, it is unlikely that
pollution by accreted terrestrial material is a viable explanation for
the currently observed host star metallicity trend. This also implies that
pollution studies should be undertaken for A-type and massive F-type stars as
they are more likely to display the preferential enrichment in Ti, Al and Ca
that appears to be indicative of terrestrial planet formation. \\

\acknowledgments J. C. Bond and D. P. O'Brien were funded by grant NNX09AB91G from NASA's Origins of Solar Systems Program. J. C. Bond and D. S. Lauretta were funded by grant NNX07AF96G from NASA's Cosmochemistry Program. Thanks to the anonymous reviewer for their helpful comments in the production of this paper.
\appendix

\section{Online Material: Tables}
\clearpage
\begin{deluxetable}{cccccccccccccccc}
\tabletypesize{\footnotesize}
\tablecolumns{16}
\tablewidth{0pt}
\rotate
\tablecaption{Predicted bulk elemental abundances for all simulated extrasolar terrestrial planets. All values are in wt\% of the final predicted planet for all seven sets of disk conditions examined. Planet number increases with increasing distance from the host star. \label{all_summ}}
\tablehead{\colhead{System}           & \colhead{H}           & \colhead{Mg}      &
\colhead{O}          & \colhead{S}  &
\colhead{Fe}          & \colhead{Al}    &
\colhead{Ca}  & \colhead{Na}  &
\colhead{Ni} & \colhead{Cr}& \colhead{P}  &
\colhead{Ti}          & \colhead{Si}    &
\colhead{C}}
\startdata
\multicolumn{16}{c}{55Cnc}\\																															 
\multicolumn{16}{c}{t=2.5$\times$10$^{5}$ years}\\ \hline																															 
55Cnc 1-4	&	0.00	&	16.59	&	26.02	&	2.58	&	20.92	&	1.40	&	0.65	&	0.51	&	1.27	&	0.24	&	0.16	&	 0.06	&	11.54	&	0.00	&	18.06	 \\
	&		&		&		&		&		&		&		&		&		&		&		&		&		&		&		 \\
55Cnc 2-4	&	0.00	&	16.50	&	25.88	&	2.93	&	20.79	&	1.39	&	0.65	&	0.51	&	1.26	&	0.24	&	0.16	&	 0.06	&	11.47	&	0.00	&	18.18	 \\
	&		&		&		&		&		&		&		&		&		&		&		&		&		&		&		 \\
55Cnc 3-4	&	0.00	&	16.80	&	26.21	&	2.20	&	21.14	&	1.41	&	0.71	&	0.52	&	1.28	&	0.24	&	0.16	&	 0.06	&	11.73	&	0.00	&	17.53	 \\
55Cnc 3-5	&	3.71	&	12.59	&	53.11	&	3.09	&	15.88	&	1.06	&	0.50	&	0.00	&	0.96	&	0.18	&	0.12	&	 0.05	&	8.76	&	0.00	&	0.00	 \\
\enddata
\end{deluxetable}

\clearpage
\section{Online Material: Chemical Condensation Sequence Plots}

\begin{figure}
\begin{center}
\includegraphics[width=150mm]{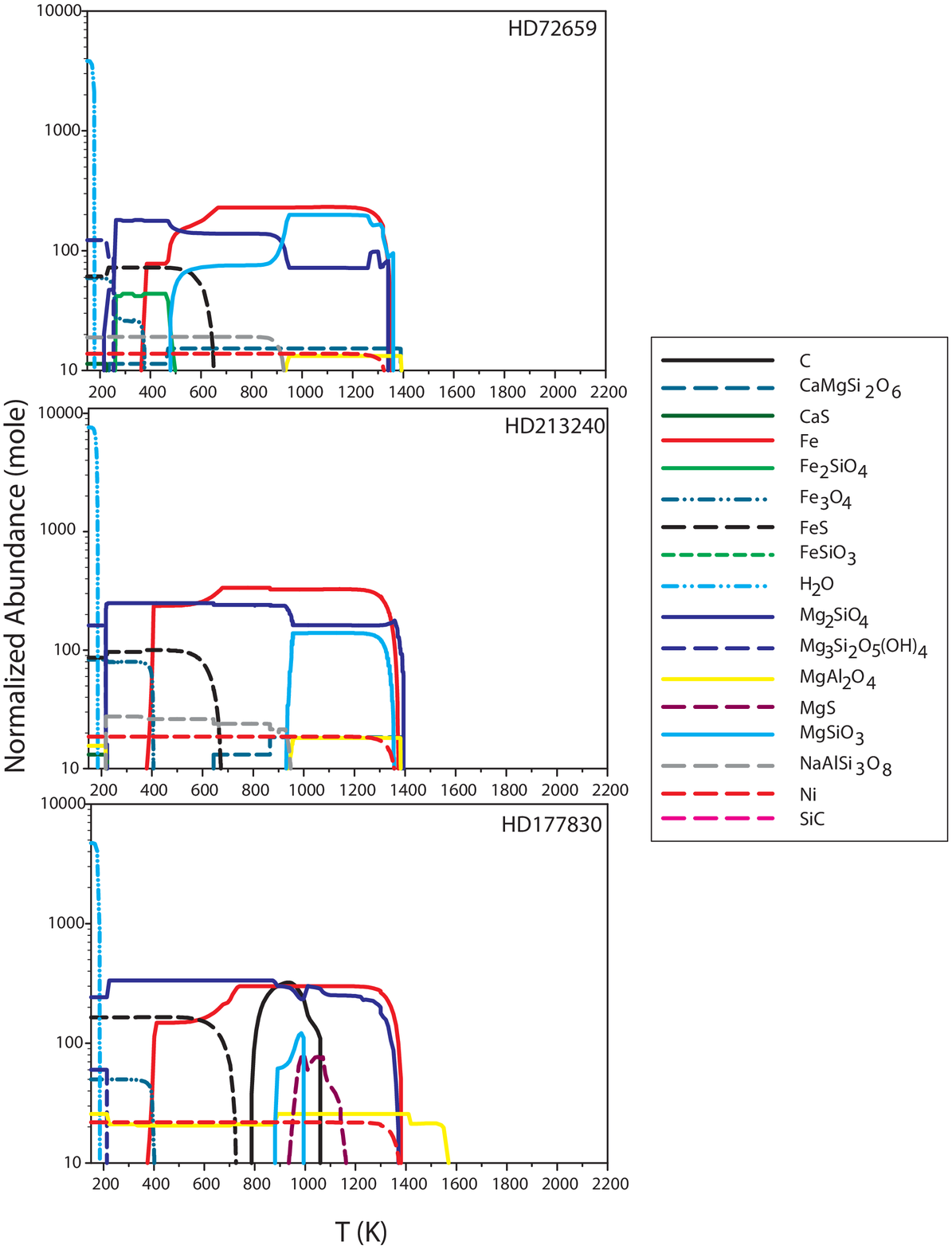}\caption{Schematic of the output obtained from HSC Chemistry for HD72659 (top), HD213240 (middle) and HD177830 (bottom) at a pressure of 10$^{-4}$ bar. Only solid species present within the system are shown. All abundances are normalized to the least abundant
species present. Input elemental abundances are shown in Table \ref{inputs}.\label{hscsumm1}}
\end{center}
\end{figure}

\clearpage

\begin{figure}
\begin{center}
\includegraphics[width=150mm]{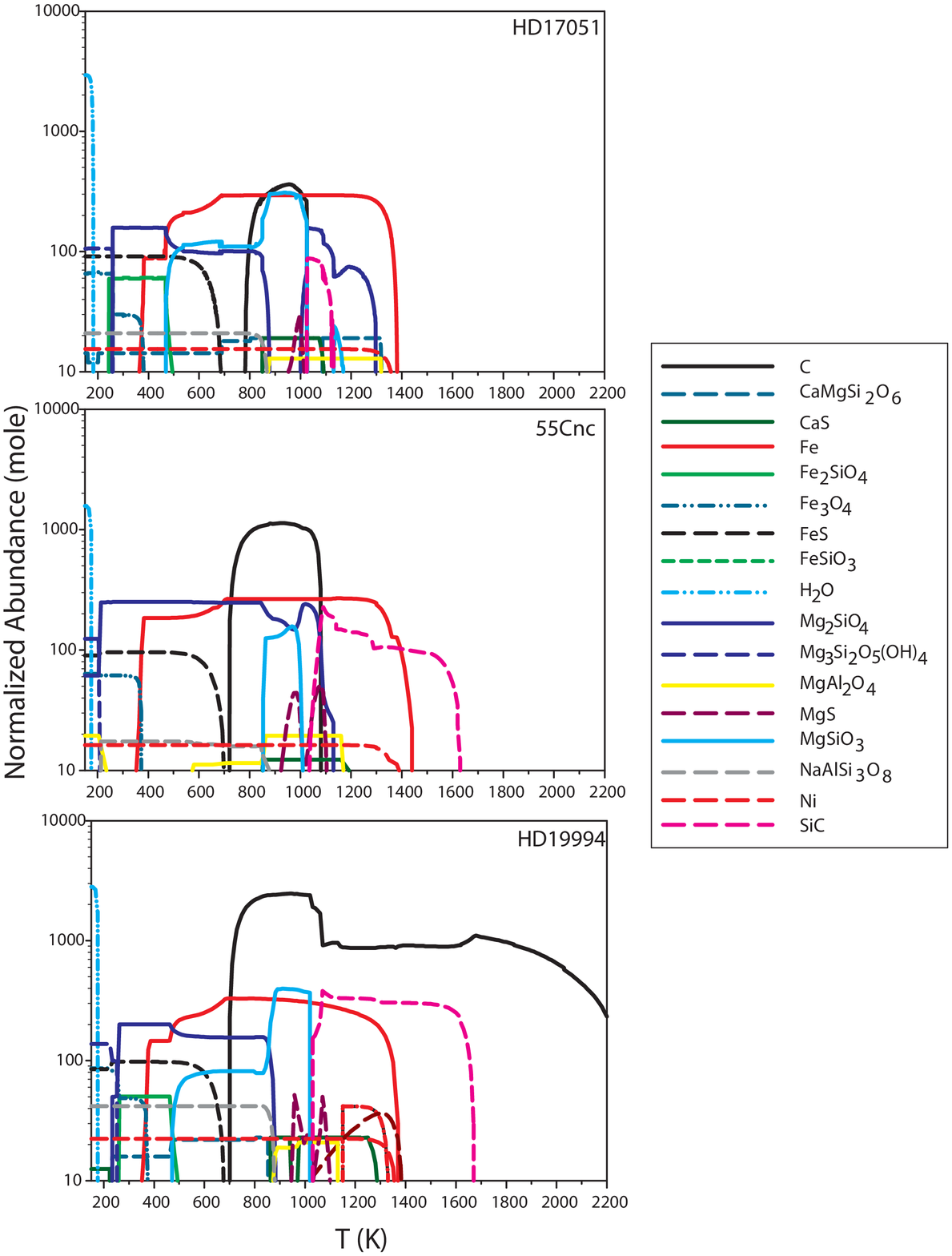}\caption{Schematic of the output obtained from HSC Chemistry for HD17051 (top), HD55Cnc (middle) and HD19994 (bottom)
at a pressure of 10$^{-4}$ bar. Only solid species present within the system are shown. All abundances are normalized to the least abundant
species present. Input elemental abundances are shown in Table \ref{inputs}.\label{hscsumm2}}
\end{center}
\end{figure}

\clearpage
\begin{figure}
\begin{center}
\includegraphics[width=150mm]{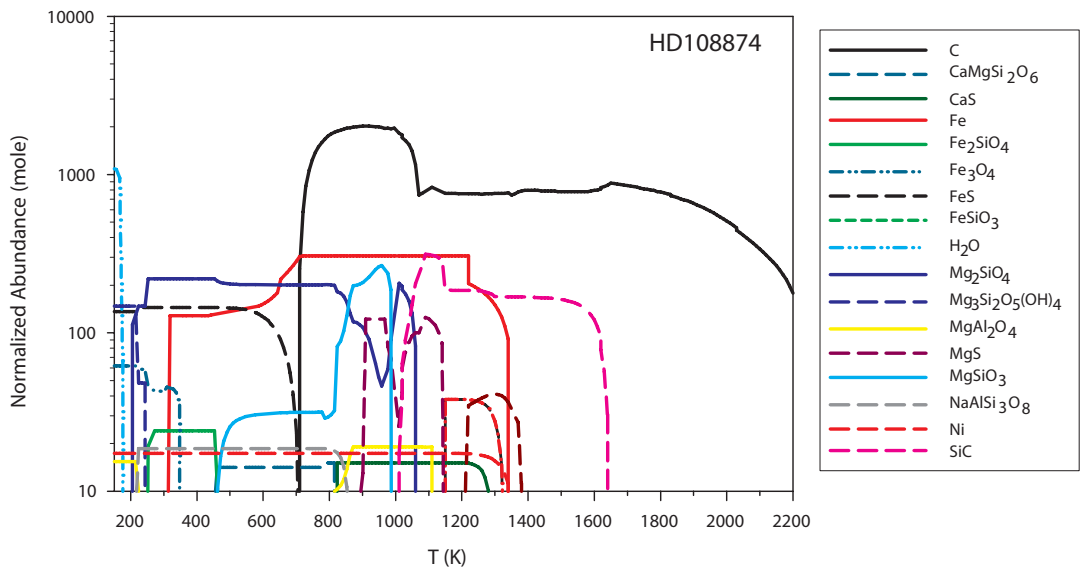}\caption{Schematic of the output obtained from HSC Chemistry for HD108874
at a pressure of 10$^{-4}$ bar. Only solid species present within the system are shown. All abundances are normalized to the least abundant
species present. Input elemental abundances are shown in Table \ref{inputs}.\label{hscsumm3}}
\end{center}
\end{figure}

\clearpage

\bibliography{references}
\clearpage

\begin{figure}
\begin{center}
\includegraphics[width=130mm]{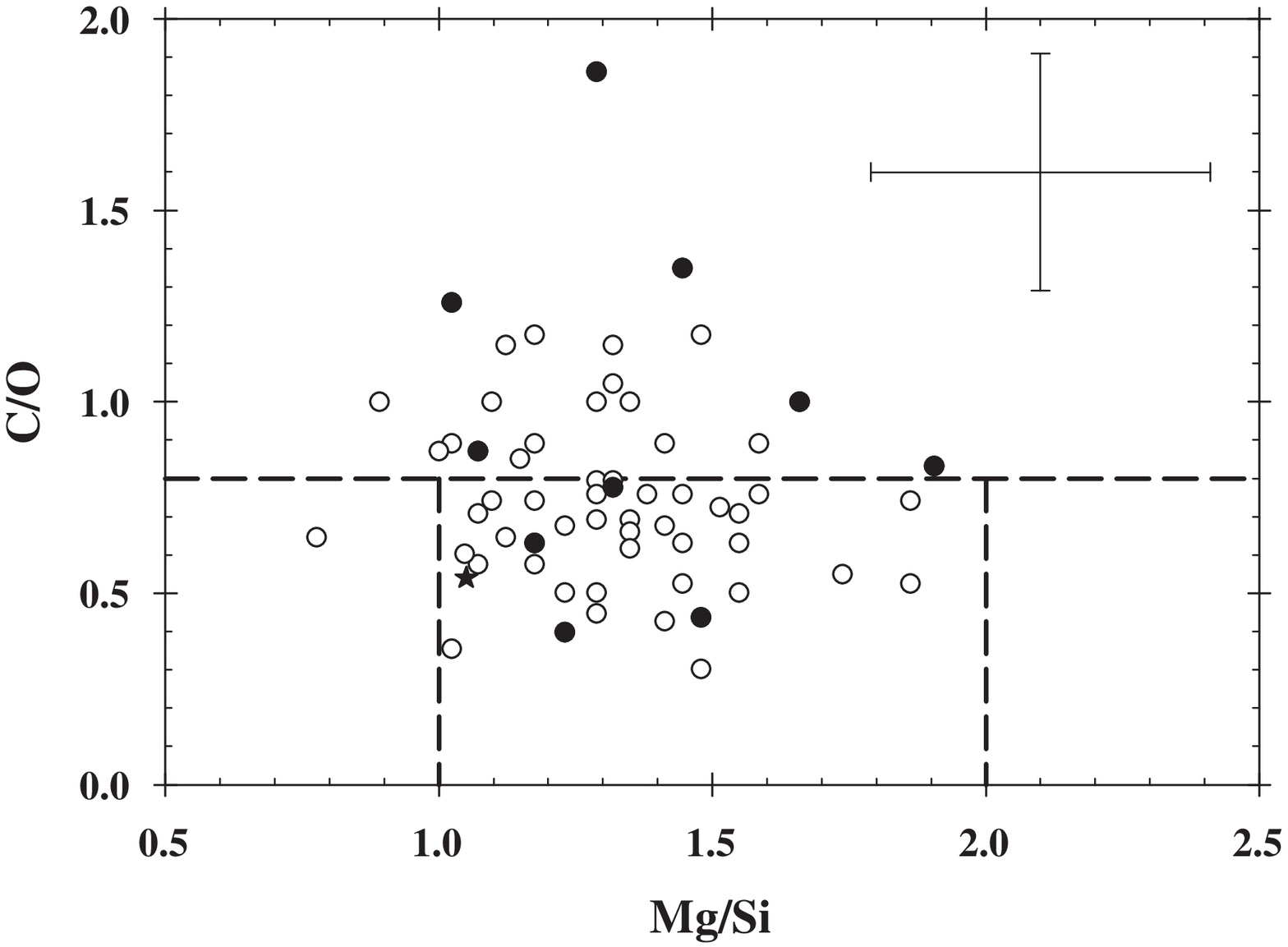} \caption[Mg/Si vs. C/O for known planetary host stars.]{Mg/Si vs. C/O for known planetary host stars
with reliable stellar abundances. Filled circles represent those systems selected for this study. Stellar photospheric values were taken from \cite{gilli} (Si, Mg), \cite{be} (Mg), \cite{eca} (C) and
\cite{oxygen} (O). Solar values are shown by the black star and were taken from \cite{asp}. The dashed line indicates a C/O value of 0.8 and
marks the transitions between a silicate-dominated composition and a carbide-dominated composition at 10$^{-4}$ bar. Average 2-$\sigma$ error bars
shown in upper right. All ratios are elemental number ratios, \emph{not} solar normalized logarithmic values.\label{regions}}
\end{center}
\end{figure}

\clearpage

\begin{figure}
\begin{center}
\includegraphics[width=150mm]{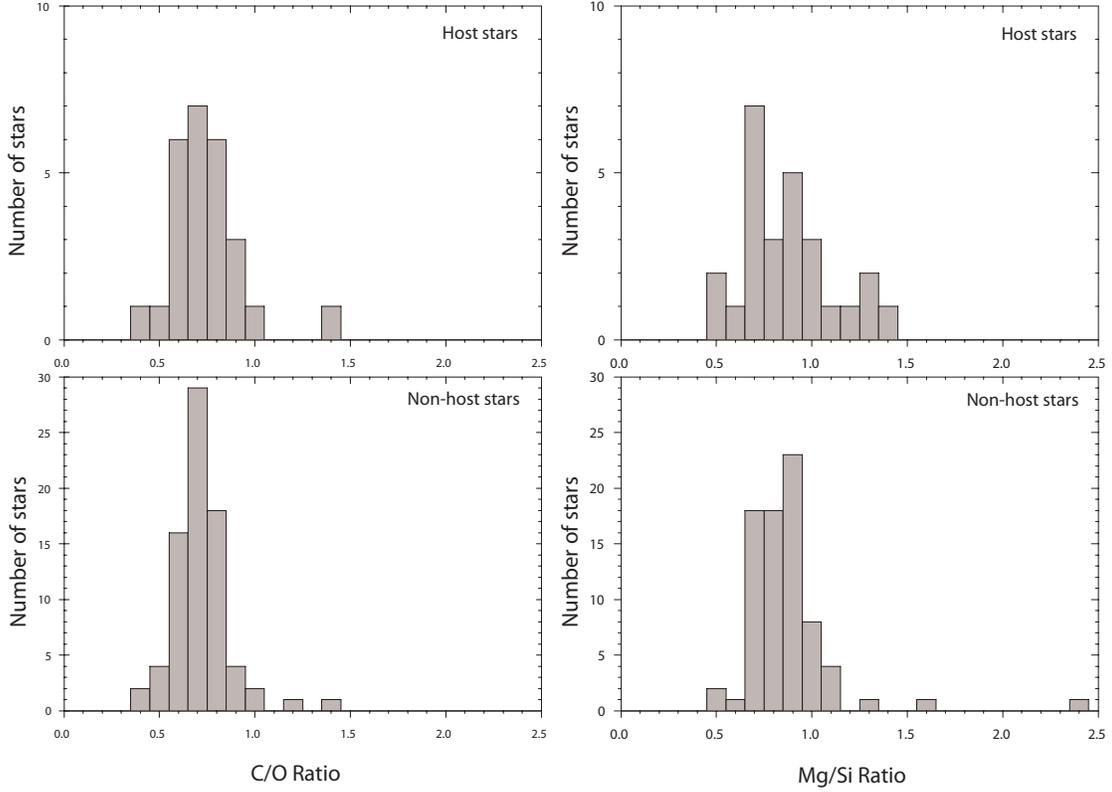} \caption[C/O and Mg/Si distributions for host and non-host stars]{C/O and Mg/Si distributions for host and
non-host stars based on the abundances determined in \cite{bond:2008}. \emph{Left:} C/O distributions for host (top) and
non-host (bottom) stars. \emph{Right:} Mg/Si distributions for host (top) and non-host (bottom) stars. All ratios are elemental number ratios, \emph{not} solar normalized logarithmic values. Note that these values are for a different dataset to that shown in Figure \ref{regions} and utilized in this work.\label{sicomp1}}
\end{center}
\end{figure}

\clearpage

\begin{figure}
\begin{center}
\includegraphics[height=80mm, width=150mm]{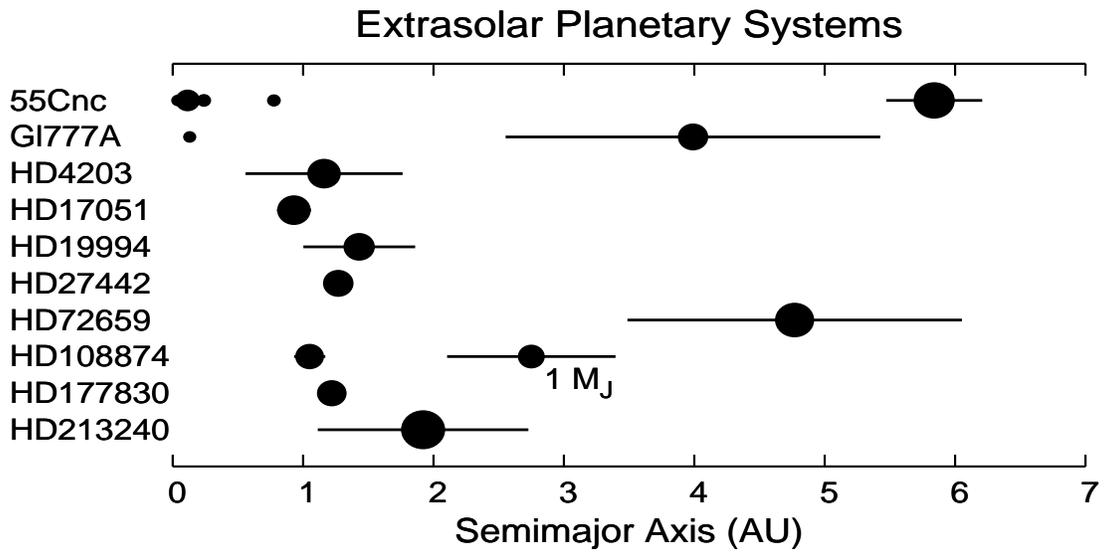} \caption[Schematic of the Extrasolar Planetary Systems studied.]{Location of known giant planets in the
systems selected for study. The horizontal lines indicate the variation from periastron and apastron. The size of the circles scales with the
planetary Msin$\textit{i}$ value. All planets are assumed to have zero inclination. All values taken from the \cite{but-cat}
catalog.\label{all_sys}}
\end{center}
\end{figure}

\clearpage

\begin{figure}
\begin{center}
\includegraphics[height=86mm, width=140mm]{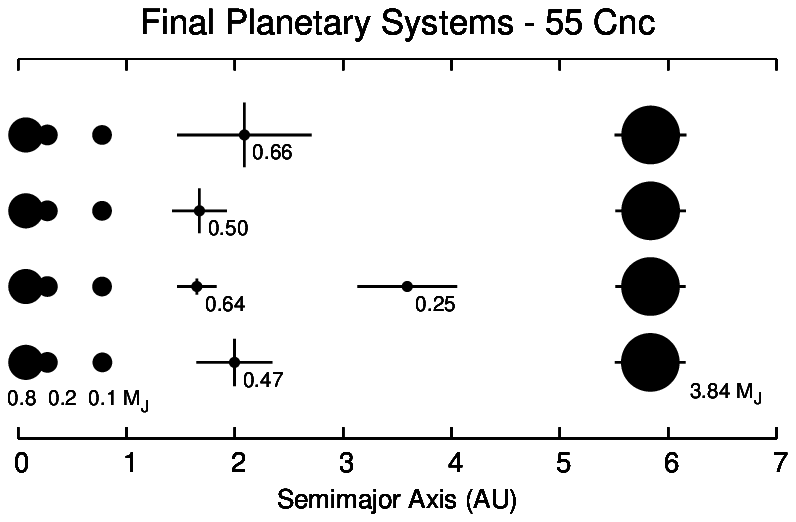} \includegraphics[height=86mm, width=140mm]{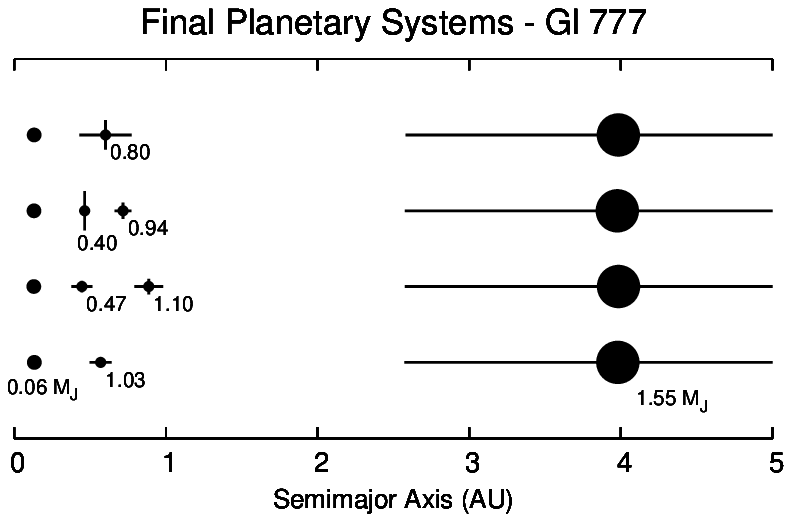} \caption[Schematic of the results of the dynamical simulations for 55Cancri and Gl777.]
{Schematic of the results of the dynamical simulations for 55Cancri (top panel) and Gl777 (bottom panel). Known giant planets are also shown
with their masses in Jupiter masses (M$_{J}$). The horizontal lines indicate the range in distance from apastron to periastron. The vertical
lines indicate variation in height above the midplane due to orbital inclination. Numerical values represent the mass of the planet in Earth
masses.\label{results1}}
\end{center}
\end{figure}

\clearpage

\begin{figure}
\begin{center}
\includegraphics[height=86mm, width=140mm]{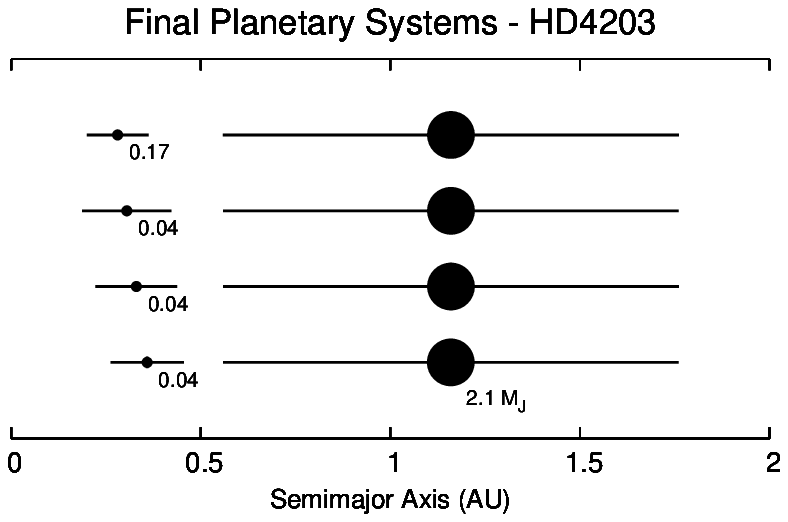} \includegraphics[height=86mm, width=140mm]{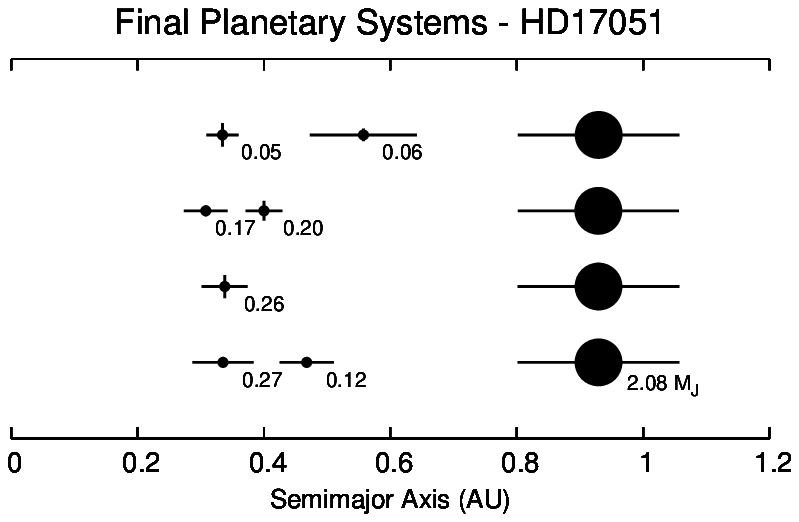} \caption[Schematic of the results of the dynamical simulations for HD4203 and HD17051.]
{Schematic of the results of the dynamical simulations for HD4203 (top panel) and HD17051 (bottom panel). Known giant planets are also shown with
their masses in Jupiter masses (M$_{J}$). The horizontal lines indicate the range in distance from apastron to periastron. The vertical lines
indicate variation in height above the midplane due to orbital inclination. Numerical values represent the mass of the planet in Earth
masses.\label{results2}}
\end{center}
\end{figure}

\clearpage

\begin{figure}
\begin{center}
\includegraphics[height=86mm, width=140mm]{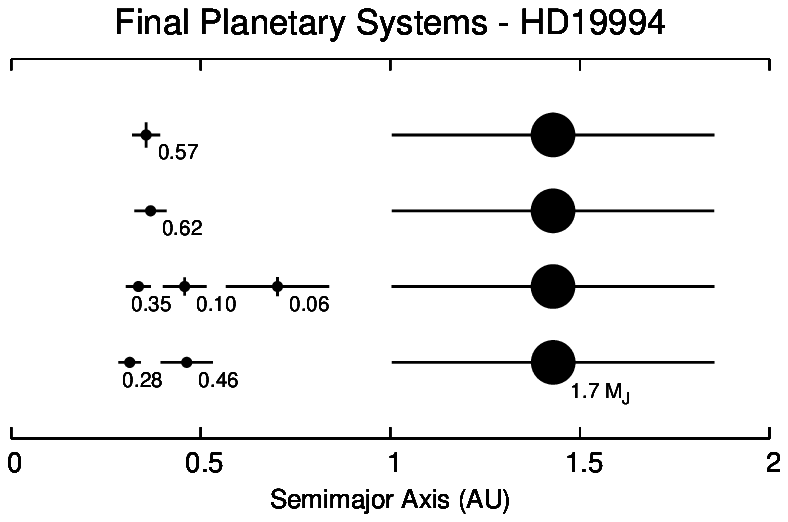} \includegraphics[height=86mm, width=140mm]{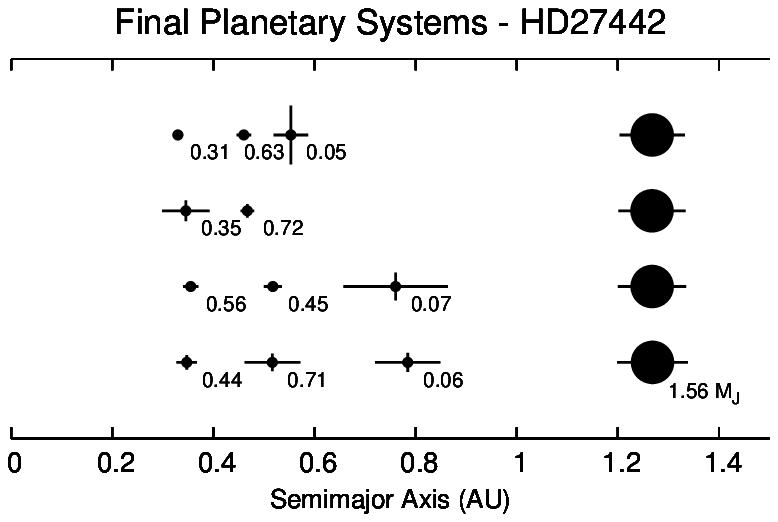} \caption[Schematic of the results of the dynamical simulations for HD19994 and HD27442.]
{Schematic of the results of the dynamical simulations for HD19994 (top panel) and HD27442 (bottom panel). Known giant planets are also shown
with their masses in Jupiter masses (M$_{J}$). The horizontal lines indicate the range in distance from apastron to periastron. The vertical
lines indicate variation in height above the midplane due to orbital inclination. Numerical values represent the mass of the planet in Earth
masses.\label{results3}}
\end{center}
\end{figure}

\clearpage

\begin{figure}
\begin{center}
\includegraphics[height=86mm, width=140mm]{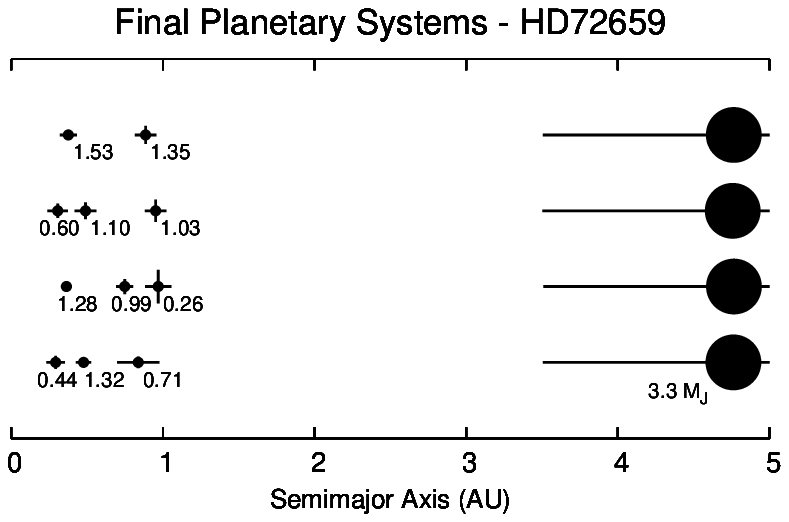} \includegraphics[height=86mm, width=140mm]{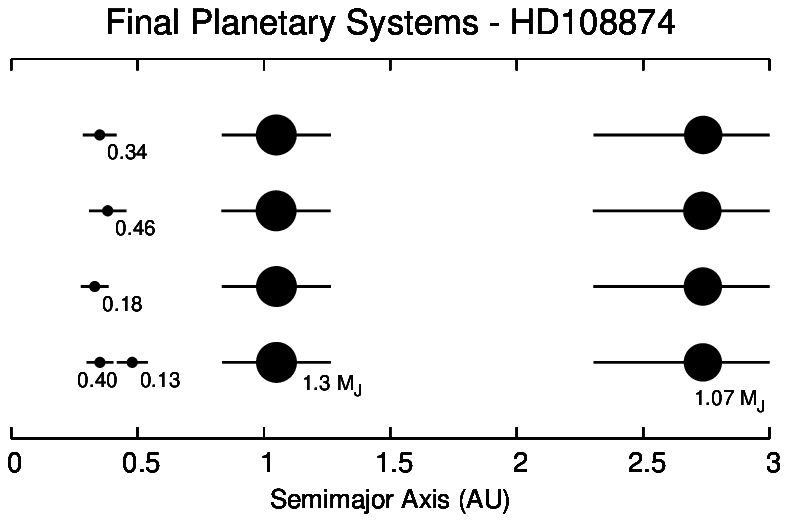} \caption[Schematic of the results of the dynamical simulations for HD72659 and HD108874.]
{Schematic of the results of the dynamical simulations for HD72659 (top panel) and HD108874 (bottom panel). Known giant planets are also shown
with their masses in Jupiter masses (M$_{J}$). The horizontal lines indicate the range in distance from apastron to periastron. The vertical
lines indicate variation in height above the midplane due to orbital inclination. Numerical values represent the mass of the planet in Earth
masses.\label{results4}}
\end{center}
\end{figure}

\clearpage

\begin{figure}
\begin{center}
\includegraphics[height=86mm, width=140mm]{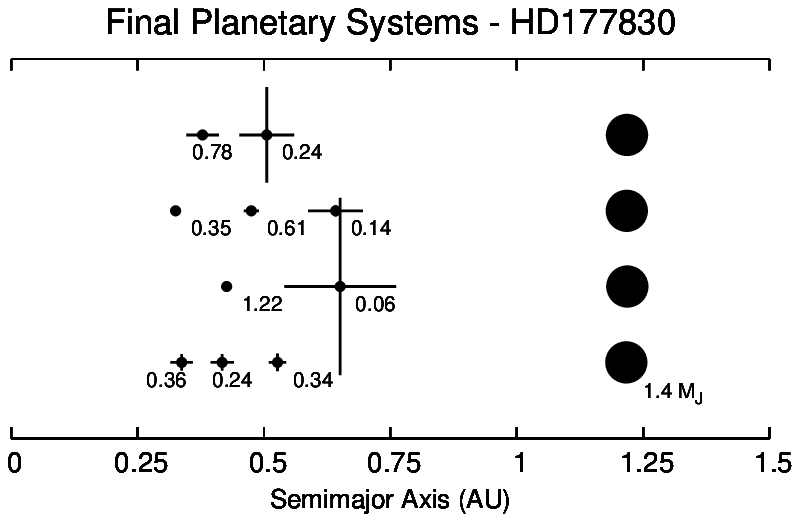} \includegraphics[height=86mm, width=140mm]{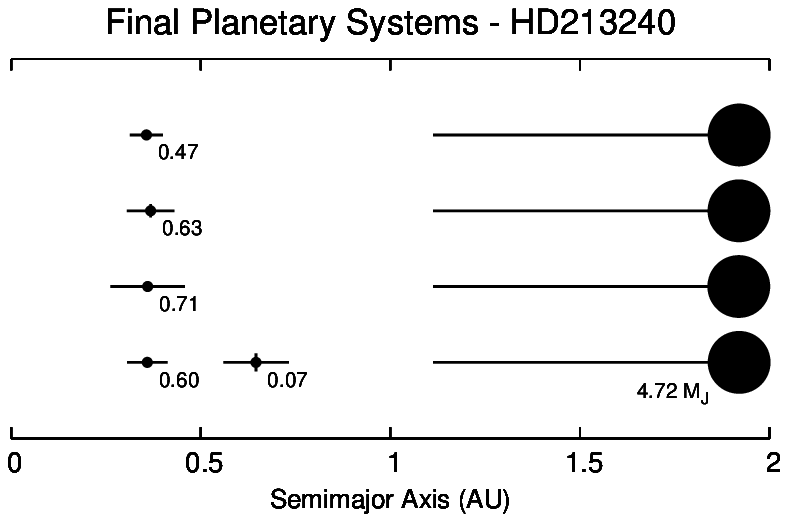} \caption[Schematic of the results of the dynamical simulations for HD177830 and HD213240.]
{Schematic of the results of the dynamical simulations for HD177830 (top panel) and HD213240 (bottom panel). Known giant planets are also shown with their masses in Jupiter masses
(M$_{J}$). The horizontal lines indicate the range in distance from apastron to periastron. The vertical lines indicate variation in height
above the midplane due to orbital inclination. Numerical values represent the mass of the planet in Earth masses.\label{results5}}
\end{center}
\end{figure}

\clearpage

\begin{figure}
\begin{center}
\includegraphics[width=130mm]{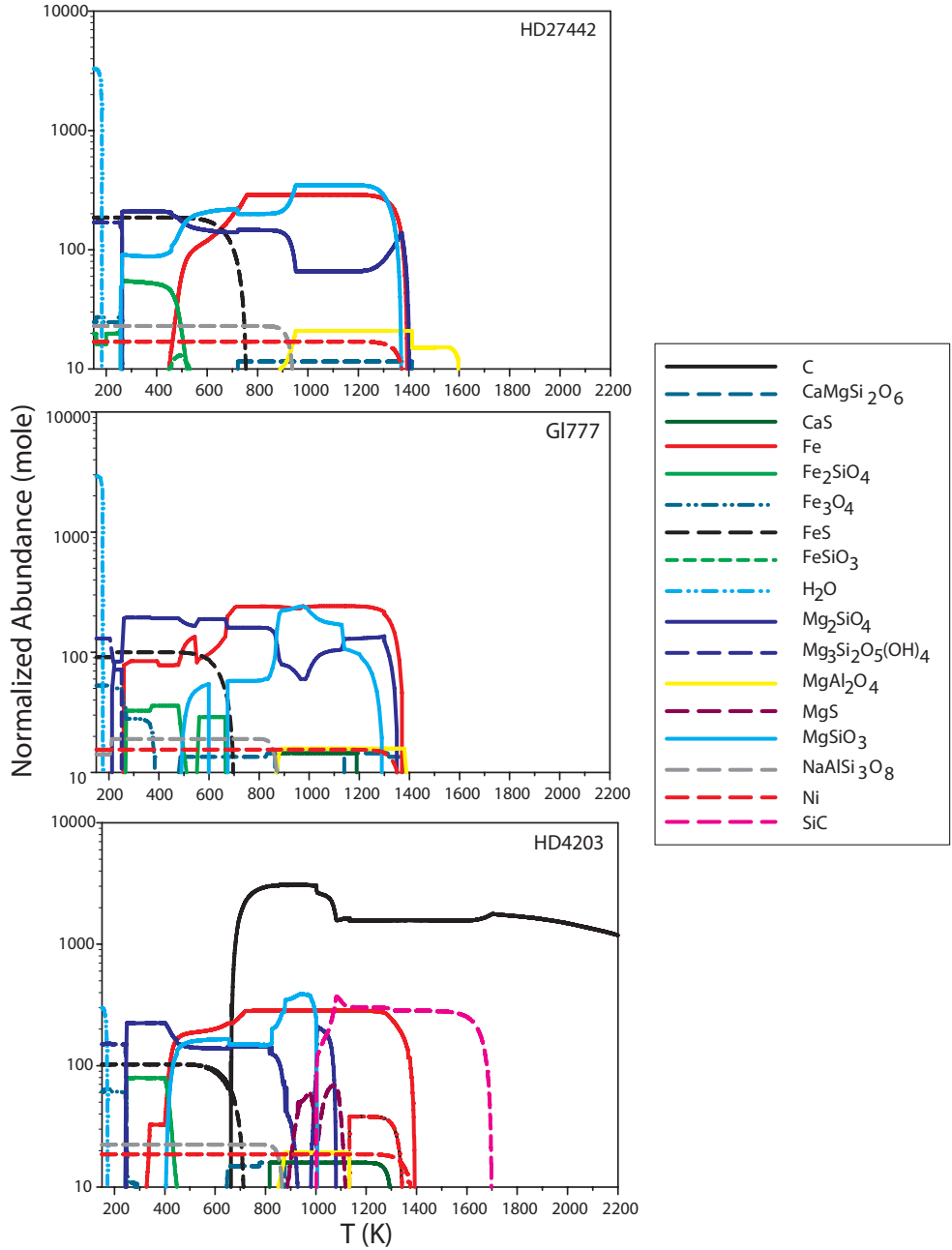}\caption[HSC Chemistry output]{Schematic of the output obtained from HSC Chemistry for 3 representative systems: HD27442 (top), Gl777 (middle) and HD4203 (bottom). All simulations were run with a system pressure of 10$^{-4}$ bar. Only solid species present within the system are shown. All abundances are normalized to the least abundant species present. Input elemental abundances are shown in Table \ref{inputs}.Note that although Gl777 is described as a low-C enriched systems, C and other carbide phases only appear for pressures at and below 10$^{-5}$ bar and are thus absent from the current figure.\label{combchem}}
\end{center}
\end{figure}

\clearpage

\begin{figure}
\begin{center}
\includegraphics[width=150mm]{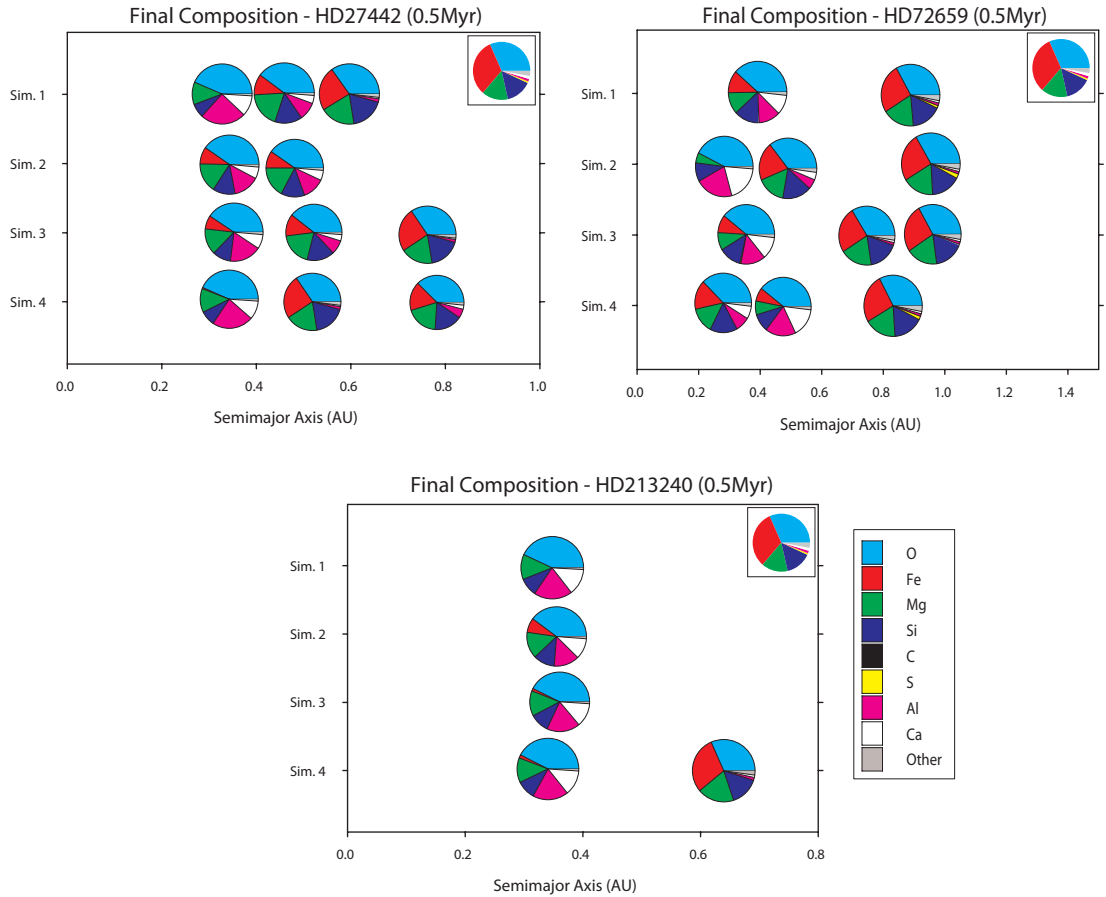}\caption[Schematic of the planetary abundances for HD27442, HD72659 and HD213240.]
{Schematic of the bulk elemental planetary composition for the Earth-like planetary systems HD27442 (top left), HD72659 (top right) and HD213240
(bottom). All values are wt\% of the final simulated planet. Values are shown for the terrestrial planets produced in each of the four
simulations run for the system. Size of bodies is not to scale. Earth values taken from \cite{kandl} are shown in the upper right of each panel for comparison. \label{earthlike}}
\newpage
\end{center}
\end{figure}

\clearpage

\begin{figure}
\begin{center}
\includegraphics[width=150mm]{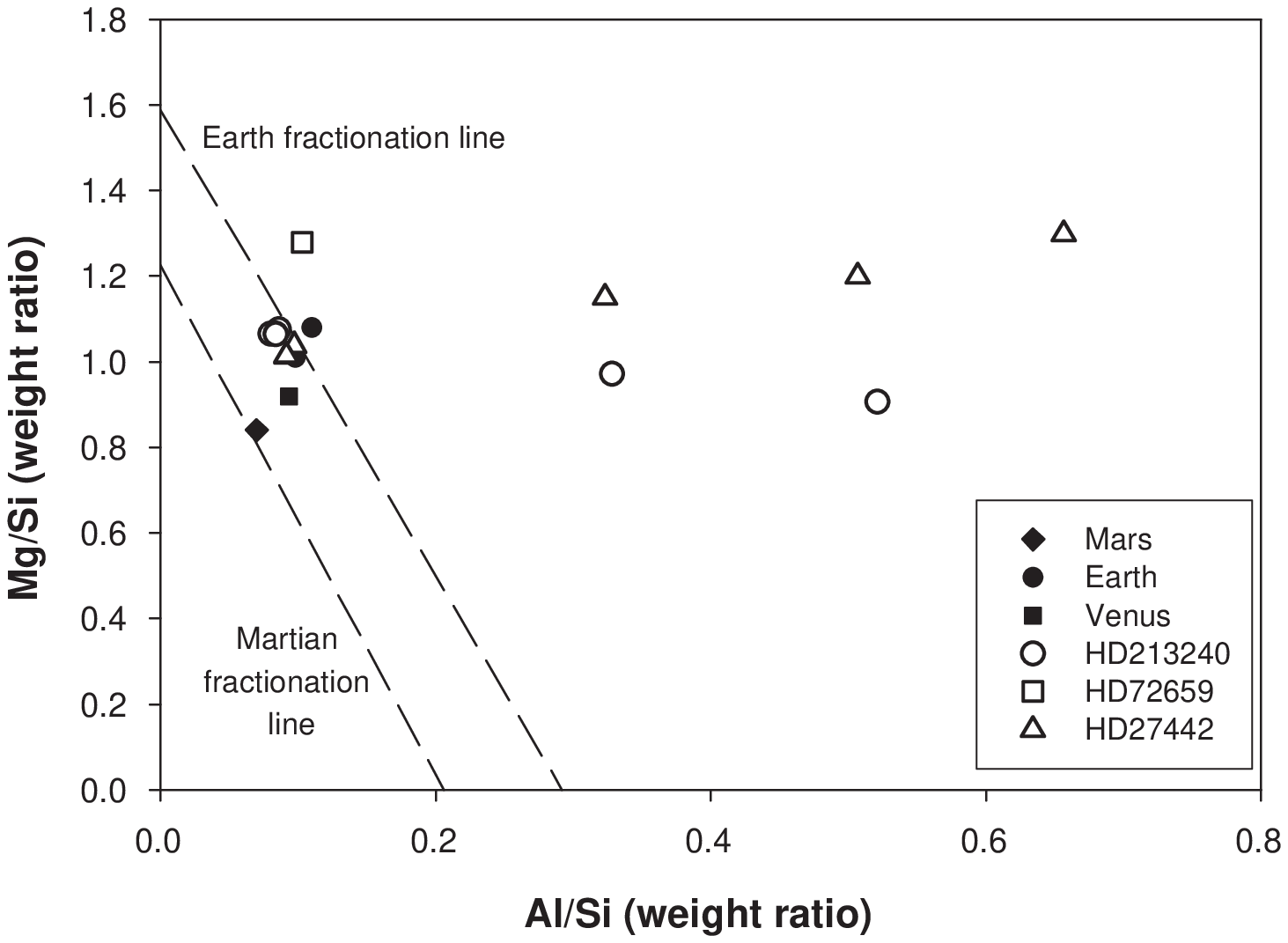}\caption[Al/Si v. Mg/Si for the planets of HD27442, HD72659 and HD213240.]
{Al/Si v. Mg/Si for the planets of HD27442 (triangles), HD72659 (squares) and HD213240 (circles). Values are for disk conditions at 5$\times$10$^{5}$ years. Earth values are shown as filled circles and are taken from \cite{kandl} and \cite{mands}. Martian values are shown as filled diamonds and are taken from \cite{landf}. Venus values are shown as filled squares and are taken from \cite{manda}. Note that the values for HD27442, HD72659 and HD213240 all extend off to the right, reaching Mg/Si values of up to 3.5.\label{EGPratios}}
\newpage
\end{center}
\end{figure}

\clearpage

\begin{figure}
\begin{center}
\includegraphics[width=150mm]{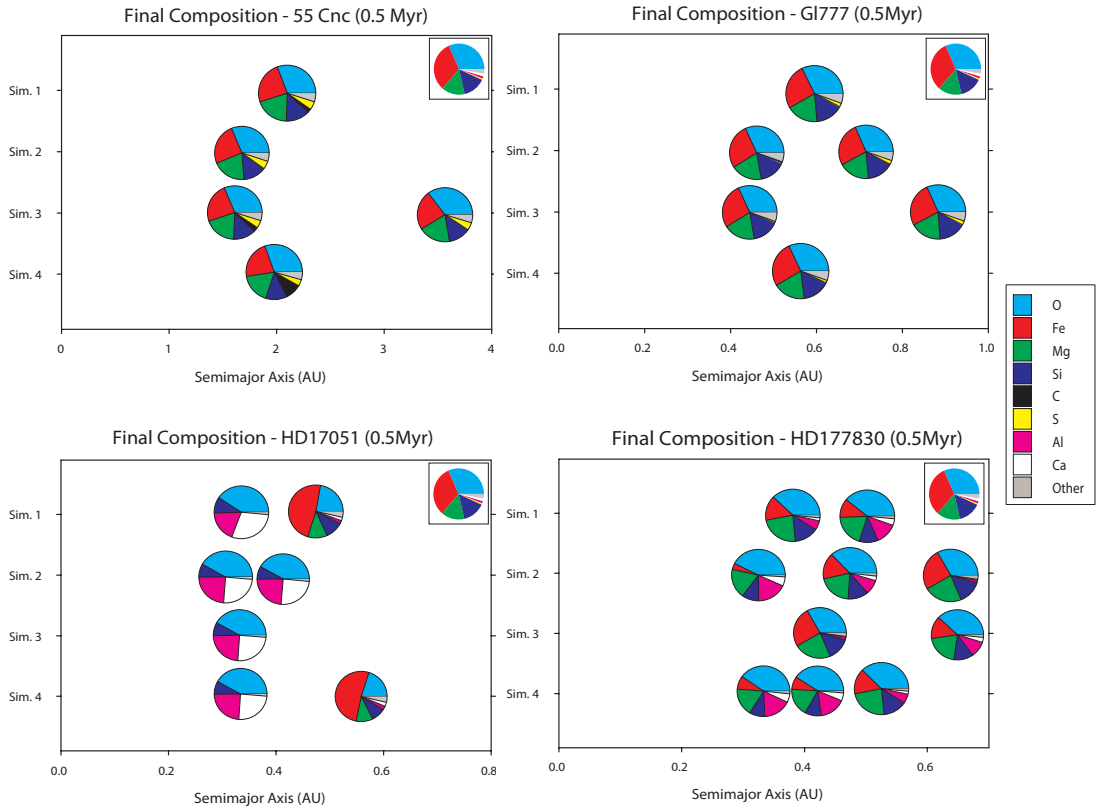}\caption[Schematic of the planetary abundances for low C planets.]
{Schematic of the bulk elemental planetary composition for the low C-enrichment systems 55Cnc (top left), Gl777 (top right), HD17051 (bottom
left) and HD177830 (bottom right). All values are wt\% of the final simulated planet. Values are shown for the terrestrial planets produced in
each of the four simulations run for the system. Size of bodies is not to scale. Earth values taken from \cite{kandl} are shown in the upper right of each panel for comparison.\label{lowC}}
\newpage
\end{center}
\end{figure}

\clearpage

\begin{figure}
\begin{center}
\includegraphics[width=150mm]{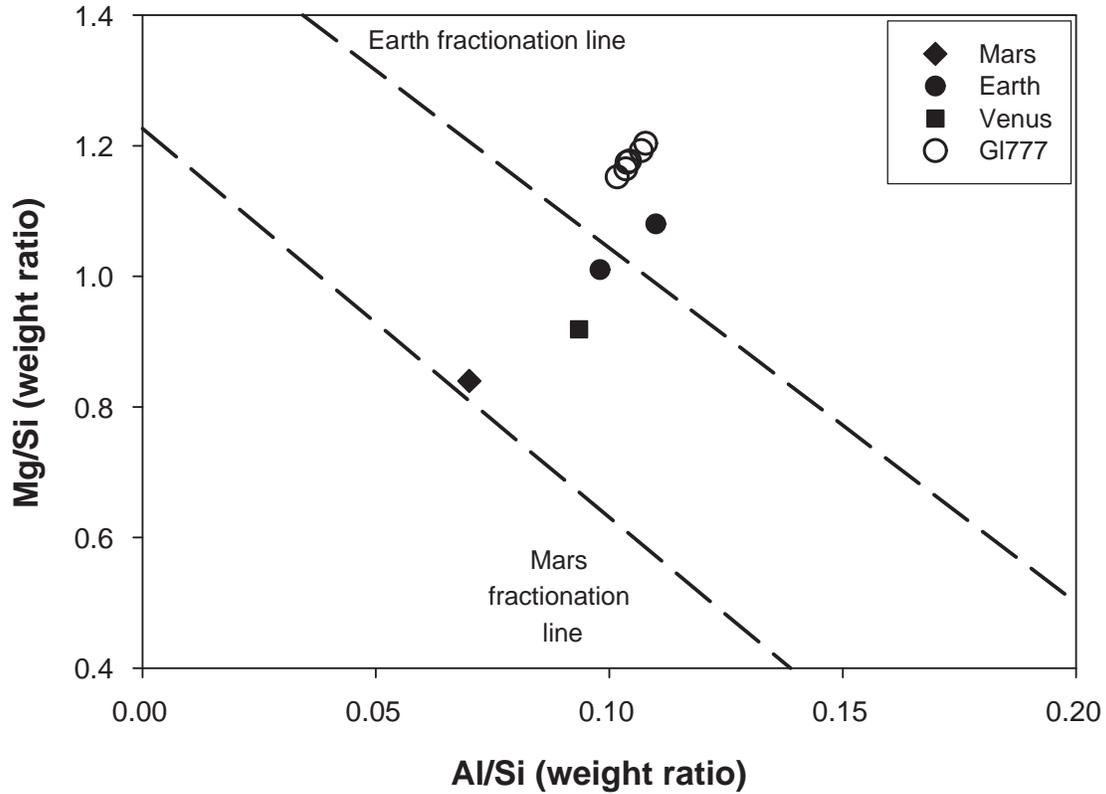}\caption[Al/Si v. Mg/Si for planets of Gl777.]
{Al/Si v. Mg/Si for planets of Gl777. Open circles
indicate values for simulated terrestrial planets produced with disk conditions at t = 5$\times$10$^{5}$ years. Values at all other times are concentrated at the 5$\times$10$^{5}$ years
values and omitted for clarity. Earth values are shown as filled circles and are taken from \cite{kandl} and \cite{mands}. Martian values are shown as filled diamonds and are taken from \cite{landf}. Venus values are shown as filled squares and are taken from \cite{manda}.\label{Gl777ratios}}
\newpage
\end{center}
\end{figure}

\clearpage

\begin{figure}
\begin{center}
\includegraphics[width=150mm]{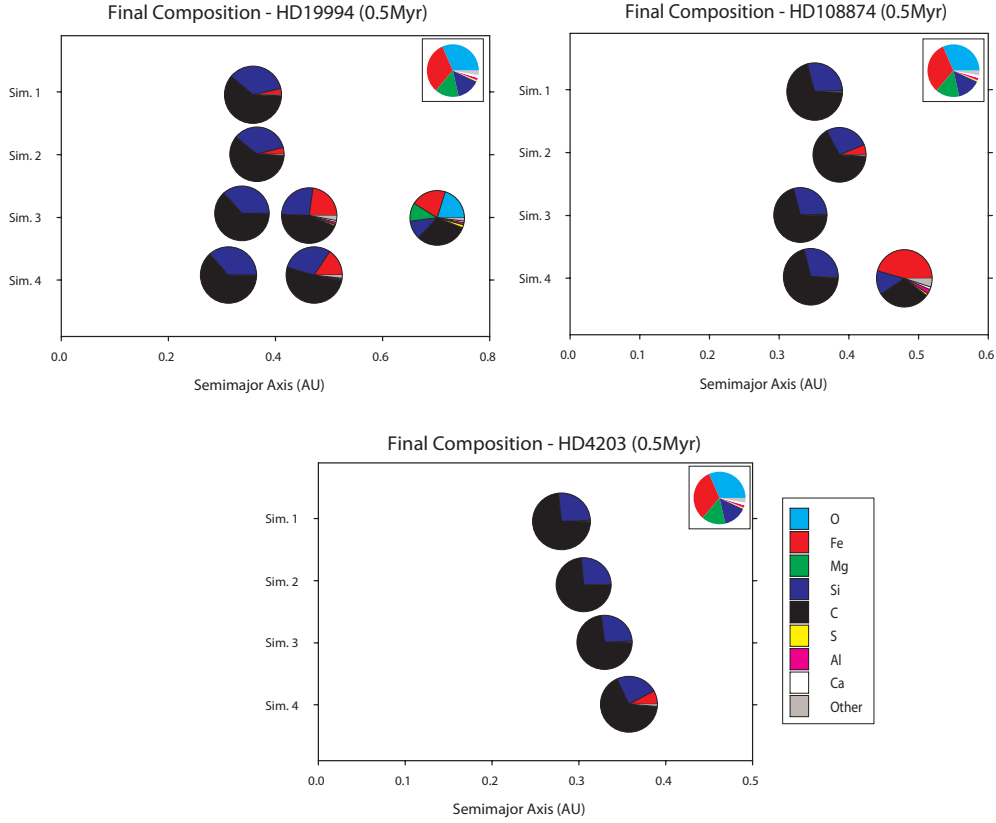}\caption[Schematic of the planetary abundances for HD19994, HD108874 and HD4203.]
{Schematic of the bulk elemental planetary composition for the high C-enrichment systems HD19994 (top left), HD108874 (top right) and HD4203
(bottom). All values are wt\% of the final simulated planet. Values are shown for the terrestrial planets produced in each of the four
simulations run for the system. Size of bodies is not to scale. Earth values taken from \cite{kandl} are shown in the upper right of each panel for comparison.\label{highC}}
\newpage
\end{center}
\end{figure}

\clearpage

\begin{figure}
\begin{center}
\includegraphics[height=89.9mm, width=150mm]{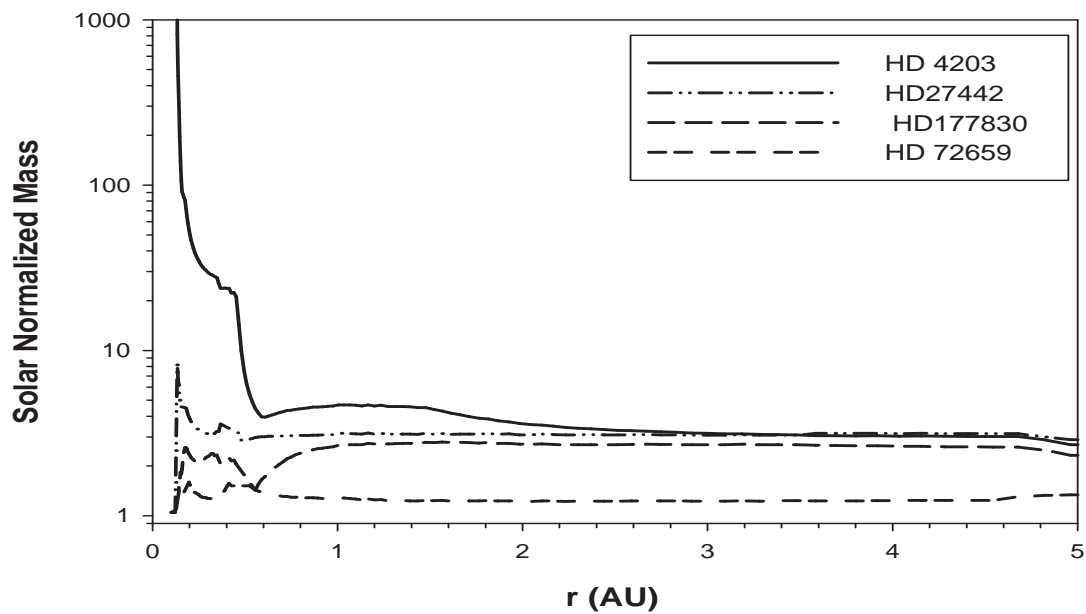} \caption
{Solid mass distribution within the disk for four known extrasolar planetary systems. All distributions are normalized to the Solar distribution. Mass distributions are shown for HD4203 (solid) (Mg/Si= 1.29, C/O=1.86), HD27442 (dashed-dotted) (Mg/Si= 1.17, C/O=0.63), HD177830 (long dash) (Mg/Si= 1.91, C/O=0.83) and HD72659 (short dash) (Mg/Si= 1.23, C/O=0.40). \label{massdist}}
\end{center}
\end{figure}

\clearpage

\begin{figure}
\begin{center}
\includegraphics[width=150mm]{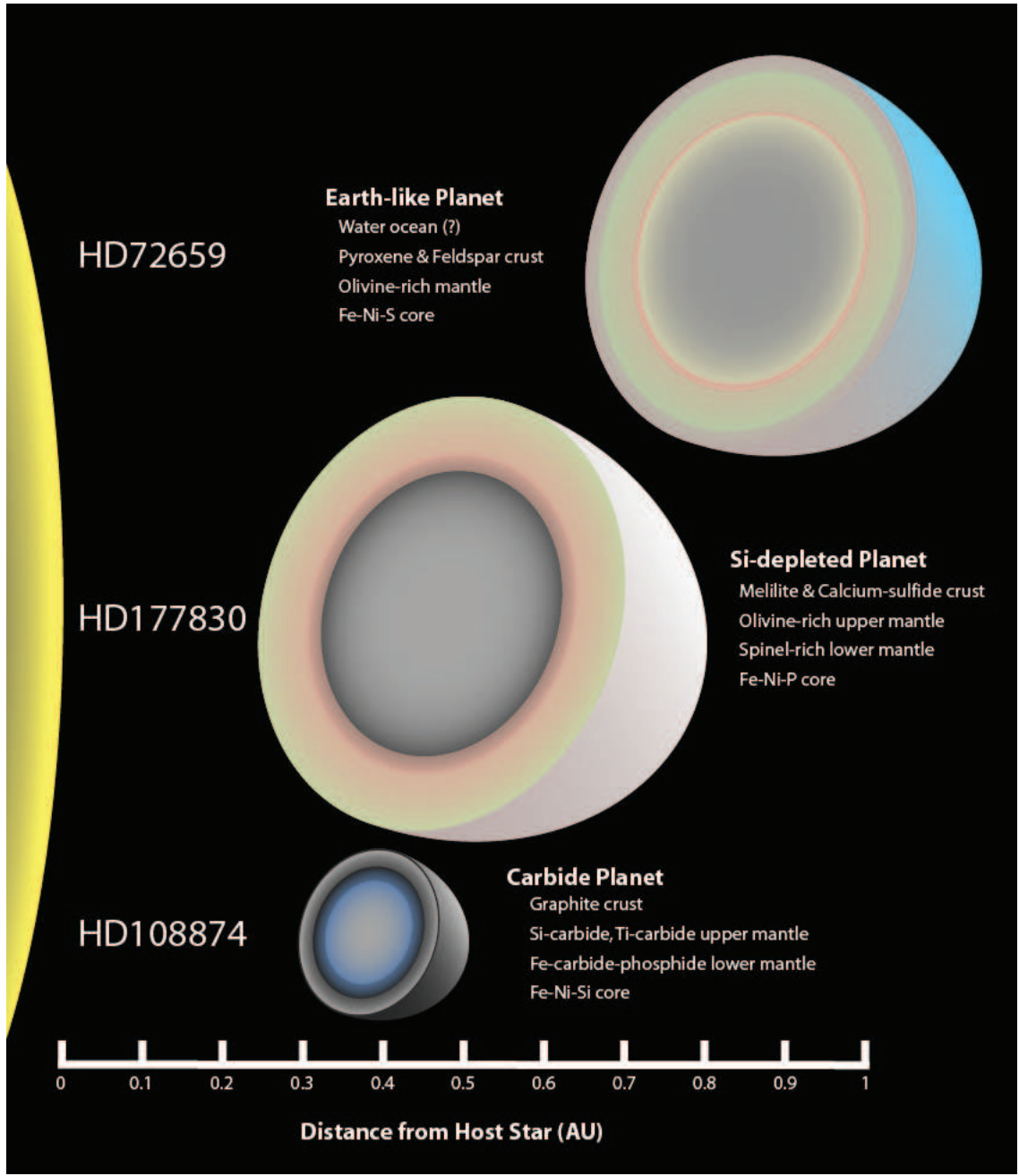}\caption[Schematic of terrestrial planet interiors for Gl777.]
{Schematic of notional interior models are based on calculations of bulk planetary compositions for disk conditions at t =
5$\times$10$^{5}$ years resulting from three different planetary systems: Gl777 (HD190360) (top), HD177830 (middle) and HD108874 (bottom). Figures are to scale for planet and layer sizes and planet location.\label{interior}}
\newpage
\end{center}
\end{figure}

\clearpage






\clearpage

\begin{table}
\begin{center}
\caption[Statistical analysis of the host and non-host star distributions of Mg/Si and C/O.]{Statistical analysis of the host and non-host star
distributions of Mg/Si and C/O. All values are based on the abundances determined in \cite{bond:2008}. The quoted uncertainty is the standard error in
the mean. All ratios are elemental number ratios, \emph{not} solar normalized logarithmic values. \label{compsi}} \vspace{0.1in}
\begin{tabular}{cccc}
\hline \hline
& Mean & Median & Standard \\
&&&Deviation \\
\hline
Mg/Si:&&& \\
Host Stars & 0.83$\pm$ 0.04 & 0.80&0.22\\
Non-Host Stars & 0.80$\pm$ 0.03 & 0.79&0.16\\
&&&\\
C/O:&&& \\
Host Stars & 0.67$\pm$ 0.03 & 0.68&0.23\\
Non-Host Stars & 0.67$\pm$ 0.03 & 0.69&0.23\\
\hline
\end{tabular}
\end{center}
\end{table}

\clearpage

\begin{table}
\begin{center}
\caption[Orbital parameters of known extrasolar planets]{Orbital parameters of known extrasolar planets for the systems studied. Values taken
from the University of California catalog located at www.exoplanets.org\label{EGPprop}} 
\begin{tabular}{cccc}
\hline Planet    &  M &   a & e     \\
   &   (M$_{\rm Jupiter}$)  & (AU) & \\
\hline \hline
55Cnc-b &   0.82    &   0.11    &   0.02    \\
55Cnc-c &   0.17    &   0.24    &   0.05    \\
55Cnc-d &   3.84    &   5.84    &   0.08    \\
55Cnc-e &   0.02    &   0.04    &   0.09    \\
55Cnc-f &   0.14    &   0.70    &   0.20    \\
    &       &       &       \\
Gl777-b &   1.55    &   4.02    &   0.35    \\
Gl777-c &   0.06    &   0.13    &   0.07    \\
    &       &       &       \\
HD4203-b    &   2.07    &   1.16    &   0.52    \\
    &       &       &       \\
HD17051-b    &   2.08    &   0.93    &   0.14    \\
    &       &       &       \\
HD19994-b   &   1.69    &   1.43    &   0.30    \\
    &       &       &       \\
HD27442-b   &   1.56    &   1.27    &   0.06    \\
    &       &       &       \\
HD72659-b   &   3.30    &   4.76    &   0.26    \\
    &       &       &       \\
HD108874-b  &   1.30    &   1.05    &   0.21    \\
HD108874-c  &   1.07    &   2.75    &   0.16    \\
    &       &       &       \\
HD177830-b  &   1.43    &   1.22    &   0.03    \\
    &       &       &       \\
HD213240-b  &   4.72    &   1.92    &   0.42    \\
    &       &       &       \\

\hline
\end{tabular}
\end{center}
\end{table}

\clearpage

\begin{deluxetable}{ccccccccccc}
\tabletypesize{\footnotesize} \tablecolumns{11} \tablewidth{0pt} \rotate \tablecaption{Target star elemental abundances in standard logarithmic
units, normalized to H and Solar values. A $-$ indicates that a value was not available for a given element or star. See text for
references.\label{EGPchem}} \tablehead{\colhead{} & \colhead{55Cnc} & \colhead{Gl777A}      & \colhead{HD4203}          & \colhead{HD17051}  &
\colhead{HD19994}          & \colhead{HD27442}    & \colhead{HD72659} & \colhead{HD108874}  & \colhead{HD177830} & \colhead{HD213240}}
\startdata
\multicolumn{1}{c}{ [Fe/H]  }   &   0.33    &   0.24    &   0.40    &   0.24    &   0.03    &   0.23    &   0.33    &   0.26    &   0.17    &   0.39    \\
\multicolumn{1}{c}{ [C/H]   }   &   0.31    &   0.29    &   0.45    &   0.39    &   $-$0.11 &   0.21    &   0.48    &   0.28    &   0.17    &   0.34    \\
\multicolumn{1}{c}{ [O/H]   }   &   0.13    &   0.22    &   0.00    &   0.11    &   0.11    &   $-$0.10  &   0.38    &   0.16    &   0.35    &   0.36    \\
\multicolumn{1}{c}{ [Na/H]  }   &   0.26    &   0.26    &   0.42    &   0.48    &   0.07    &   0.13    &   0.37    &   0.24    &   0.22    &   0.41    \\
\multicolumn{1}{c}{ [Mg/H]  }   &   0.48    &   0.33    &   0.48    &   0.21    &   0.11    &   0.27    &   0.56    &   0.19    &   0.23    &   0.51    \\
\multicolumn{1}{c}{ [Al/H]  }   &   0.47    &   0.34    &   0.51    &   0.32    &   0.07    &   0.30    &   0.54    &   0.19    &   0.20    &   0.53    \\
\multicolumn{1}{c}{ [Si/H]  }   &   0.29    &   0.24    &   0.44    &   0.23    &   0.05    &   0.14    &   0.31    &   0.19    &   0.09    &   0.47    \\
\multicolumn{1}{c}{ [S/H]   }   &   0.12    &   0.10    &   0.20    &   $-$0.05 &   $-$0.23 &    $-$   &   $-$    &   0.00    &   $-$0.10  &   $-$    \\
\enddata
\end{deluxetable}

\clearpage

\begin{deluxetable}{crrrrrrrrrrr}
\tabletypesize{\scriptsize} \tablecolumns{11} \tablewidth{0pt} \rotate \tablecaption{Target star elemental abundances as number of atoms and
normalized to 10$^{6}$Si atoms. A $-$ indicates that a value was not available for a given element or star. See text for
references.\label{EGPchemSi}} \tablehead{\colhead{Element}           &\colhead{Solar}           & \colhead{55Cnc} & \colhead{Gl777A}      &
\colhead{HD4203}          & \colhead{HD17051}  & \colhead{HD19994}          & \colhead{HD27442}    & \colhead{HD72659} & \colhead{HD108874}  &
\colhead{HD177830} & \colhead{HD213240}} \startdata

Fe  &   8.32    $\times$10$^{5}$    &   9.12    $\times$10$^{5}$    &   8.32    $\times$10$^{5}$    &   7.59    $\times$10$^{5}$    &   9.77    $\times$10$^{5}$    &   8.51    $\times$10$^{5}$    &   6.92    $\times$10$^{5}$    &   7.94    $\times$10$^{5}$    &   10.23   $\times$10$^{6}$    &   8.71    $\times$10$^{5}$    &   10.00   $\times$10$^{6}$    \\
C   &   1.02    $\times$10$^{7}$    &   1.07    $\times$10$^{7}$    &   1.15    $\times$10$^{7}$    &   1.05    $\times$10$^{7}$    &   1.26    $\times$10$^{7}$    &   1.48    $\times$10$^{7}$    &   7.59    $\times$10$^{6}$    &   7.08    $\times$10$^{6}$    &   1.20    $\times$10$^{7}$    &   1.51    $\times$10$^{7}$    &   1.23    $\times$10$^{7}$    \\
O   &   1.55    $\times$10$^{7}$    &   1.07    $\times$10$^{7}$    &   1.48    $\times$10$^{7}$    &   5.62    $\times$10$^{6}$    &   1.45    $\times$10$^{7}$    &   1.17    $\times$10$^{7}$    &   1.20    $\times$10$^{7}$    &   1.78    $\times$10$^{7}$    &   8.91    $\times$10$^{6}$    &   1.82    $\times$10$^{7}$    &   2.82    $\times$10$^{7}$    \\
Na  &   6.03    $\times$10$^{4}$    &   5.62    $\times$10$^{4}$    &   6.31    $\times$10$^{4}$    &   5.75    $\times$10$^{4}$    &   6.76    $\times$10$^{4}$    &   1.07    $\times$10$^{5}$    &   5.25    $\times$10$^{4}$    &   6.31    $\times$10$^{4}$    &   5.89    $\times$10$^{4}$    &   6.92    $\times$10$^{4}$    &   8.13    $\times$10$^{4}$    \\
Mg  &   1.07    $\times$10$^{6}$    &   1.66    $\times$10$^{6}$    &   1.32    $\times$10$^{6}$    &   1.17    $\times$10$^{6}$    &   1.07    $\times$10$^{6}$    &   1.02    $\times$10$^{6}$    &   1.17    $\times$10$^{6}$    &   1.23    $\times$10$^{6}$    &   1.45    $\times$10$^{6}$    &   1.91    $\times$10$^{6}$    &   1.48    $\times$10$^{6}$    \\
\enddata
\end{deluxetable}

\clearpage

\begin{deluxetable}{cccc}
\tabletypesize{\footnotesize}
\tablecolumns{11}
\tablewidth{0pt}
\tablecaption{Chemical species included in the equilibrium calculations of HSC
Chemistry. \label{HSC}}
\tablehead{\multicolumn{4}{c}{\textbf{Gaseous Species}}}
\startdata
 Al     &       CrO     &       MgOH        &       PN   \\
 AlH        &       CrOH        &       MgS     &       PO   \\
 AlO        &       CrS     &       N       &       PS   \\
 Al$_{2}$O      &       Fe      &       N$_{2}$     &       S    \\
 AlOH       &       FeH     &       NH$_{3}$        &       S$_{2}$  \\
 AlS        &       FeO     &       NO      &       SN   \\
 C      &       FeOH        &       NS      &       SO   \\
 CH$_{4}$       &       FeS     &       Na      &       SO$_{2}$ \\
\enddata
\end{deluxetable}

\clearpage

\begin{deluxetable}{crrrrrrrrrr}
\tabletypesize{\footnotesize}
\tablecolumns{11}
\tablewidth{0pt}
\rotate
\tablecaption{HSC Chemistry input values for the extrasolar planetary systems
studied. All inputs are entered into the simulations of HSC Chemistry their elemental and gaseous state. All values are in moles and are based
on the stellar abundance values listed in Tables \ref{EGPchem} and \ref{EGPchemSi}. \label{inputs}}
\tablehead{\colhead{Element}           & \multicolumn{10}{c}{System}           \\
\colhead{}           &
\colhead{55Cnc}           & \colhead{Gl777A}      &
\colhead{HD4203}          & \colhead{HD17051}  &
\colhead{HD19994}          & \colhead{HD27442}    &
\colhead{HD72659}  & \colhead{HD108874}  &
\colhead{HD177830} & \colhead{HD213240}}
\startdata
H	&	1.0	$\times$10$^{	12	}$&	1.0	$\times$10$^{	12	}$&	1.0	$\times$10$^{	12	}$&	1.0	$\times$10$^{	12	}$&	1.0	$\times$10$^{	12	}$&	 1.0	$\times$10$^{	12	}$&	1.0	$\times$10$^{	12	}$&	1.0	$\times$10$^{	12	}$&	1.0	$\times$10$^{	12	}$&	1.0	$\times$10$^{	12	}$\\
He	&	8.5	$\times$10$^{	10	}$&	8.5	$\times$10$^{	10	}$&	8.5	$\times$10$^{	10	}$&	8.5	$\times$10$^{	10	}$&	8.5	$\times$10$^{	10	}$&	 8.5	$\times$10$^{	10	}$&	8.5	$\times$10$^{	10	}$&	8.5	$\times$10$^{	10	}$&	8.5	$\times$10$^{	10	}$&	8.5	$\times$10$^{	10	}$\\
C	&	7.4	$\times$10$^{	8	}$&	7.1	$\times$10$^{	8	}$&	1.0	$\times$10$^{	9	}$&	6.9	$\times$10$^{	8	}$&	8.9	$\times$10$^{	8	}$&	 7.9	$\times$10$^{	8	}$&	2.8	$\times$10$^{	8	}$&	5.9	$\times$10$^{	8	}$&	1.1	$\times$10$^{	9	}$&	5.4	$\times$10$^{	8	}$\\
N	&	1.8	$\times$10$^{	8	}$&	1.7	$\times$10$^{	8	}$&	2.5	$\times$10$^{	8	}$&	1.7	$\times$10$^{	8	}$&	2.2	$\times$10$^{	8	}$&	 2.0	$\times$10$^{	8	}$&	0.7	$\times$10$^{	8	}$&	1.5	$\times$10$^{	8	}$&	2.7	$\times$10$^{	8	}$&	1.3	$\times$10$^{	8	}$\\
O	&	7.4	$\times$10$^{	8	}$&	9.1	$\times$10$^{	8	}$&	5.5	$\times$10$^{	8	}$&	7.9	$\times$10$^{	8	}$&	7.1	$\times$10$^{	8	}$&	 1.3	$\times$10$^{	9	}$&	7.1	$\times$10$^{	8	}$&	4.4	$\times$10$^{	8	}$&	1.3	$\times$10$^{	9	}$&	1.2	$\times$10$^{	9	}$\\
Na	&	3.9	$\times$10$^{	6	}$&	3.9	$\times$10$^{	6	}$&	5.6	$\times$10$^{	6	}$&	3.7	$\times$10$^{	6	}$&	6.5	$\times$10$^{	6	}$&	 5.5	$\times$10$^{	6	}$&	2.5	$\times$10$^{	6	}$&	2.9	$\times$10$^{	6	}$&	5.0	$\times$10$^{	6	}$&	3.6	$\times$10$^{	6	}$\\
Mg	&	1.2	$\times$10$^{	8	}$&	8.1	$\times$10$^{	7	}$&	1.2	$\times$10$^{	8	}$&	5.9	$\times$10$^{	7	}$&	6.2	$\times$10$^{	7	}$&	 1.2	$\times$10$^{	8	}$&	4.9	$\times$10$^{	7	}$&	7.1	$\times$10$^{	7	}$&	1.4	$\times$10$^{	8	}$&	6.5	$\times$10$^{	7	}$\\
Al	&	8.7	$\times$10$^{	6	}$&	6.5	$\times$10$^{	6	}$&	9.6	$\times$10$^{	6	}$&	4.6	$\times$10$^{	6	}$&	6.2	$\times$10$^{	6	}$&	 1.0	$\times$10$^{	7	}$&	3.5	$\times$10$^{	6	}$&	5.9	$\times$10$^{	6	}$&	1.0	$\times$10$^{	7	}$&	4.7	$\times$10$^{	6	}$\\
\enddata
\end{deluxetable}

\begin{table}
\begin{center}
\caption[Extrasolar host star accretion rates]{Stellar accretion rates for the extrasolar host stars studied. Stellar masses
were obtained from the Simbad database. Solar values are the nominal model determined by \cite{hersant}. See text
for details on the scaling relations applied.\label{mdot}} \vspace{0.3in}
\begin{tabular}{ccc}
\hline System    &   Stellar Mass  &   $\dot{M}$     \\
   &   (M$_{\bigodot}$)  &   (M$_{\bigodot}$/year) \\
\hline \hline
Solar   &   1.00   &   5.00$\times$10$^{-6}$    \\
    &       &                \\
55Cnc   &   1.03   &   5.23$\times$10$^{-6}$    \\
    &       &                \\
Gl777   &   1.04   &   5.30$\times$10$^{-6}$    \\
    &       &                \\
HD4203  &   1.06   &   5.46$\times$10$^{-6}$    \\
    &       &                \\
HD17051  &   1.11   &   5.50$\times$10$^{-6}$    \\
    &       &                \\
HD19994 &   1.35   &   7.84$\times$10$^{-6}$    \\
    &       &                \\
HD27442 &   1.20   &   6.00$\times$10$^{-6}$    \\
    &       &                \\
HD72659 &   0.95   &   4.63$\times$10$^{-6}$    \\
    &       &                \\
HD108874    &   1.00   &   5.00$\times$10$^{-6}$    \\
    &       &                \\
HD177830    &   1.48   &   9.00$\times$10$^{-6}$    \\
    &       &                \\
HD213240    &   1.22   &   6.10$\times$10$^{-6}$    \\

\hline
\end{tabular}
\end{center}
\end{table}

\begin{table}
\begin{center}
\caption[Stellar convective zone masses]{Convective zone masses for each of the target stars. T$_{eff}$ values taken from \cite{ism04}. See text
for details on the determination of M$_{CZ}$. \label{MCZ}} \vspace{0.3in}
\begin{tabular}{ccc}
\hline System    &   T$_{eff}$  &   M$_{CZ}$     \\
   &   (K)  &   (M$_{\bigodot}$) \\
\hline \hline 55Cnc   &   5279    &   0.0398    \\
    &       &                \\
Gl777   &   5584    &   0.0316    \\
    &       &                \\
HD4203  &   5636    &   0.0288 \\
    &       &                \\
HD17051  &   6253    &  0.0035     \\
    &       &                \\
HD19994 &   6190    &   0.0045    \\
    &       &                \\
HD27442  &  4825     &  0.0562     \\
    &       &                \\
HD72659 &   5995    &   0.0112    \\
    &       &                \\
HD108874    &   5596    &   0.0321    \\
    &       &                \\
HD177830    &   4804    &   0.0501    \\
    &       &                \\
HD213240  &    5984   &  0.0100   \\
    &       &                \\
\hline
\end{tabular}
\end{center}
\end{table}

\begin{deluxetable}{crrrrrrrrrrr}
\tabletypesize{\footnotesize}
\tablecolumns{11}
\tablewidth{0pt}
\rotate
\tablecaption{T$_{50 \%\:condensation}$, C/O and Mg/Si  for extrasolar planetary
systems studied in order of increasing C/O value. Solar values are also shown for comparison and are taken from \cite{bond:2009}. Where less than 50\% of an element condensed within a system, the maximum amount of condensation that occurred is shown in parentheses. All values are in K. \label{EGP_Tc}}
\tablehead{\colhead{Element}           & \multicolumn{11}{c}{System}           \\
\colhead{}           &
\colhead{HD72659}           & \colhead{HD213240}      &
\colhead{HD27442}          & \colhead{Solar}  &
\colhead{Gl777A}          & \colhead{HD177830}    &
\colhead{HD17051}  & \colhead{55Cnc}  &
\colhead{HD19994} & \colhead{HD108874} & \colhead{HD4203}}
\startdata
Al	&	1665	&	1698	&	1688	&	1639	&	1636	&	1647	&	1551	&	1311	&	1320	&	1315	&	1333	 \\
C	&	$<$150	&	$<$150	&	$<$150	&	$<$150	&	$<$150	&	976 	&	955 	&	920 	&	954 	&	946	&	1049	 \\
&	&		&		&		&		&	(6\%)	&	(9\%)	&	(34\%)	&	(39\%)	&		&		 \\
Ca	&	1541	&	1562	&	1613	&	1527	&	1562	&	1556	&	1484	&	1248	&	1287	&	1289	&	1300	 \\
Cr	&	1297	&	1309	&	1327	&	1301	&	1314	&	1322	&	1316	&	1322	&	1313	&	1306	&	1324	 \\
Fe	&	1334	&	1346	&	1365	&	1339	&	1353	&	1361	&	1354	&	1361	&	1353	&	1352	&	1367	 \\
Mg	&	1359	&	1381	&	1377	&	1339	&	1328	&	1321	&	1224	&	1081	&	1075	&	1066	&	1074	 \\
Na	&	927	&	944	&	935	&	941	&	870	&	801	&	866	&	864	&	873	&	854	&	864	 \\
\enddata
\end{deluxetable}

\begin{deluxetable}{cccccccccccccccc}
\tabletypesize{\footnotesize}
\tablecolumns{16}
\tablewidth{0pt}
\rotate
\tablecaption{Predicted bulk elemental abundances for all simulated extrasolar terrestrial planets. All values are in wt\% of the final predicted planet for all seven sets of disk conditions examined. Planet number increases with increasing distance from the host star. \label{all_summ}}
\tablehead{\colhead{System}           & \colhead{H}           & \colhead{Mg}      &
\colhead{O}          & \colhead{S}  &
\colhead{Fe}          & \colhead{Al}    &
\colhead{Ca}  & \colhead{Na}  &
\colhead{Ni} & \colhead{Cr}& \colhead{P}  &
\colhead{Ti}          & \colhead{Si}    &
\colhead{C}}
\startdata
\multicolumn{16}{c}{55Cnc}\\																															 
\multicolumn{16}{c}{t=2.5$\times$10$^{5}$ years}\\ \hline																															 
55Cnc 1-4	&	0.00	&	16.59	&	26.02	&	2.58	&	20.92	&	1.40	&	0.65	&	0.51	&	1.27	&	0.24	&	0.16	&	 0.06	&	11.54	&	0.00	&	18.06	 \\
	&		&		&		&		&		&		&		&		&		&		&		&		&		&		&		 \\
55Cnc 2-4	&	0.00	&	16.50	&	25.88	&	2.93	&	20.79	&	1.39	&	0.65	&	0.51	&	1.26	&	0.24	&	0.16	&	 0.06	&	11.47	&	0.00	&	18.18	 \\
	&		&		&		&		&		&		&		&		&		&		&		&		&		&		&		 \\
55Cnc 3-4	&	0.00	&	16.80	&	26.21	&	2.20	&	21.14	&	1.41	&	0.71	&	0.52	&	1.28	&	0.24	&	0.16	&	 0.06	&	11.73	&	0.00	&	17.53	 \\
55Cnc 3-5	&	3.71	&	12.59	&	53.11	&	3.09	&	15.88	&	1.06	&	0.50	&	0.00	&	0.96	&	0.18	&	0.12	&	 0.05	&	8.76	&	0.00	&	0.00	 \\
	&		&		&		&		&		&		&		&		&		&		&		&		&		&		&		 \\
55Cnc 4-4	&	1.79	&	15.74	&	43.90	&	3.87	&	19.87	&	1.32	&	0.62	&	0.31	&	1.20	&	0.23	&	0.15	&	 0.06	&	10.95	&	0.00	&	0.00	 \\
\enddata																															
\tablecomments{Table \ref{all_summ} is published in its entirety in the
electronic edition of the {\it Astrophysical Journal}.  A portion is
shown here for guidance regarding its form and content.}
\end{deluxetable} 

\begin{deluxetable}{ccccccccccccccc}
\tabletypesize{\footnotesize}
\tablecolumns{15}
\tablewidth{0pt}
\rotate
\tablecaption{Average change in host star photospheric abundances  produced by terrestrial planet formation. Values are based on the results of four separate simulations.\label{EGPpollution}}
\tablehead{\colhead{System}           & \colhead{Mass Accreted}&\multicolumn{13}{c}{Change in Abundance}           \\
\colhead{}           &
\colhead{(M$_\bigoplus$)}           & \colhead{Mg}      &
\colhead{O}          & \colhead{S}  &
\colhead{Fe}          & \colhead{Al}    &
\colhead{Ca}  & \colhead{Na}  &
\colhead{Ni} & \colhead{Cr}& \colhead{P}  &
\colhead{Ti}          & \colhead{Si}    &
\colhead{C}}
\startdata
55Cnc	&	0.145	&	0.000	&	0.000	&	0.000	&	0.000	&	0.000	&	0.000	&	0.000	&	0.000	&	0.000	&	0.000	&	0.000	 &	0.000	&	0.000	 \\
Gl777	&	2.100	&	0.011	&	0.009	&	0.009	&	0.009	&	0.011	&	0.011	&	0.011	&	0.011	&	0.011	&	0.004	&	0.012	 &	0.011	&	0.009	 \\
HD4203	&	0.631	&	0.000	&	0.000	&	0.000	&	0.000	&	0.000	&	0.000	&	0.000	&	0.000	&	0.000	&	0.000	&	0.002	 &	0.001	&	0.001	 \\
HD17051	&	0.727	&	0.004	&	0.004	&	0.004	&	0.004	&	0.013	&	0.016	&	0.004	&	0.005	&	0.005	&	0.000	&	0.020	 &	0.005	&	0.004	 \\
HD19994	&	0.898	&	0.004	&	0.003	&	0.003	&	0.003	&	0.005	&	0.005	&	0.003	&	0.005	&	0.005	&	0.000	&	0.006	 &	0.006	&	0.005	 \\
HD27442	&	0.342	&	0.001	&	0.001	&	0.000	&	0.001	&	0.001	&	0.001	&	0.001	&	0.001	&	0.001	&	0.000	&	0.001	 &	0.001	&	0.001	 \\
HD72659	&	2.640	&	0.021	&	0.017	&	0.001	&	0.016	&	0.020	&	0.030	&	0.023	&	0.021	&	0.025	&	0.007	&	0.026	 &	0.023	&	0.016	 \\
HD108874	&	0.755	&	0.001	&	0.001	&	0.001	&	0.001	&	0.002	&	0.001	&	0.001	&	0.002	&	0.002	&	0.000	&	 0.003	&	0.002	&	0.001	 \\
HD177830	&	0.237	&	0.001	&	0.001	&	0.001	&	0.001	&	0.001	&	0.001	&	0.001	&	0.001	&	0.001	&	0.000	&	 0.001	&	0.001	&	0.001	 \\
HD213240	&	1.310	&	0.015	&	0.012	&	0.005	&	0.011	&	0.019	&	0.024	&	0.014	&	0.015	&	0.016	&	0.011	&	 0.021	&	0.014	&	0.011	 \\
\enddata
\end{deluxetable}

\end{document}